\documentclass[
reprint,
nobibnotes,
superscriptaddress,
amsmath,amssymb,
aps,
floatfix,
]{revtex4-2}

\usepackage{siunitx}
\usepackage{bm}
\usepackage[dvipsnames]{xcolor}
\usepackage{graphicx}
\usepackage[utf8]{inputenc}
\usepackage{epstopdf}
\usepackage{afterpage}
\usepackage{setspace}
\usepackage{booktabs}
\usepackage{hyperref}
\usepackage{xspace}
\usepackage[normalem]{ulem}
\usepackage{upgreek}
\usepackage{xr}

\newcommand{\RuNNer}{\textsc{RuNNer 2.0}\xspace}
\newcommand{\runnerase}{\textsc{runnerase}\xspace}

\expandafter\ifx\csname package@font\endcsname\relax\else
 \expandafter\expandafter
 \expandafter\usepackage
 \expandafter\expandafter
 \expandafter{\csname package@font\endcsname}%
\fi
\RequirePackage[normalem]{ulem}
\begin{document}

\title{\RuNNer: A Software Suite for High-Dimensional Neural Network Potentials}

\author{Alexander L. M. Knoll}
\affiliation{Lehrstuhl f\"ur Theoretische Chemie II, Ruhr-Universit\"at Bochum, 44780 Bochum, Germany}
\affiliation{Research Center Chemical Sciences and Sustainability, Research Alliance Ruhr, 44780 Bochum, Germany}
\author{Moritz R. Sch\"afer}
\affiliation{Lehrstuhl f\"ur Theoretische Chemie II, Ruhr-Universit\"at Bochum, 44780 Bochum, Germany}
\affiliation{Research Center Chemical Sciences and Sustainability, Research Alliance Ruhr, 44780 Bochum, Germany}
\author{K. Nikolas Lausch}
\affiliation{Lehrstuhl f\"ur Theoretische Chemie II, Ruhr-Universit\"at Bochum, 44780 Bochum, Germany}
\affiliation{Research Center Chemical Sciences and Sustainability, Research Alliance Ruhr, 44780 Bochum, Germany}
\author{Moritz Gubler}
\affiliation{University of Basel, Department of Physics, Klingelbergstrasse 82, CH-4056 Basel, Switzerland}
\affiliation{PSI Center for Scientific Computing, Theory and Data, Paul Scherrer Institute, 5232 Villigen PSI, Switzerland}
\author{Henry Wang}
\affiliation{Lehrstuhl f\"ur Theoretische Chemie II, Ruhr-Universit\"at Bochum, 44780 Bochum, Germany}
\affiliation{Research Center Chemical Sciences and Sustainability, Research Alliance Ruhr, 44780 Bochum, Germany}
\author{Richard Springborn}
\affiliation{Lehrstuhl f\"ur Theoretische Chemie II, Ruhr-Universit\"at Bochum, 44780 Bochum, Germany}
\affiliation{Research Center Chemical Sciences and Sustainability, Research Alliance Ruhr, 44780 Bochum, Germany}
\author{Redouan El Haouari}
\affiliation{Lehrstuhl f\"ur Theoretische Chemie II, Ruhr-Universit\"at Bochum, 44780 Bochum, Germany}
\affiliation{Research Center Chemical Sciences and Sustainability, Research Alliance Ruhr, 44780 Bochum, Germany}
\author{Alea Miako Liebetrau}
\affiliation{Lehrstuhl f\"ur Theoretische Chemie II, Ruhr-Universit\"at Bochum, 44780 Bochum, Germany}
\affiliation{Research Center Chemical Sciences and Sustainability, Research Alliance Ruhr, 44780 Bochum, Germany}
\author{Jonas A. Finkler}
\affiliation{University of Basel, Department of Physics, Klingelbergstrasse 82, CH-4056 Basel, Switzerland}
\author{Emir Kocer}
\affiliation{Lehrstuhl f\"ur Theoretische Chemie II, Ruhr-Universit\"at Bochum, 44780 Bochum, Germany}
\affiliation{Research Center Chemical Sciences and Sustainability, Research Alliance Ruhr, 44780 Bochum, Germany}
\author{Marco Eckhoff}
\affiliation{ETH Zurich, Department of Chemistry and Applied Biosciences, Vladimir-Prelog-Weg 2, 8093 Zurich, Switzerland}
\author{Stefan Goedecker}
\affiliation{University of Basel, Department of Physics, Klingelbergstrasse 82, CH-4056 Basel, Switzerland}
\author{Gunnar Schmitz}
\affiliation{Lehrstuhl f\"ur Theoretische Chemie II, Ruhr-Universit\"at Bochum, 44780 Bochum, Germany}
\affiliation{Research Center Chemical Sciences and Sustainability, Research Alliance Ruhr, 44780 Bochum, Germany}
\author{J\"{o}rg Behler}
\email{joerg.behler@rub.de}
\affiliation{Lehrstuhl f\"ur Theoretische Chemie II, Ruhr-Universit\"at Bochum, 44780 Bochum, Germany}
\affiliation{Research Center Chemical Sciences and Sustainability, Research Alliance Ruhr, 44780 Bochum, Germany}

\renewcommand{\MakeTextUppercase}{\MakeUppercase}

\begin{abstract}
We present \RuNNer, the ``Ruhr University Neural Network energy representation'', a highly optimized software suite for training and evaluating high-dimensional neural network potentials (HDNNPs) of the second, third, and fourth generation. Long-range electrostatics and charge equilibration (QEq) for the description of non-local charge transfer in fourth-generation (4G) HDNNPs are accelerated by quasi-linear-scaling plane-wave methods, reducing QEq computational complexity from $\mathcal{O}(N^3)$ to $\mathcal{O}(N\log^2 N)$ such that linear or quasi-linear scaling is achieved across all HDNNP generations. 
An optimized memory management strategy eliminates the training overhead traditionally associated with long-range interactions, allowing 4G-HDNNPs to be trained with the same efficiency as their local counterparts. Developed in modern Fortran (2003/2008 standards), combined with a hybrid MPI/OpenMP parallelization scheme, \RuNNer has been designed to run efficiently in any CPU environment, from cost-effective local workstations to massive HPC clusters. Its modular library architecture facilitates straightforward binding to external simulation software; native interfaces to LAMMPS and the Atomic Simulation Environment (ASE) provide full access to all its features, including built-in committee-based uncertainty quantification. The high efficiency and scalability of the \RuNNer ecosystem are demonstrated through detailed benchmarks.
\end{abstract}

\maketitle

\section{Introduction}\label{sec:introduction}

Machine learning potentials (MLPs) have transformed computational materials science and chemistry~\cite{P4885,P5673,P6102,P6112,P6121,P6631} by combining the computational efficiency of classical force fields with an accuracy approaching that of electronic structure methods. Two decades of methodological development have produced model architectures spanning the full range from very efficient and highly accurate potentials tailored to a particular system~\cite{P1174,P2630,P5367,P4862,P4644,P5794} to broadly applicable foundation models trained on large and diverse datasets~\cite{P7067,P6669,P7238,P7137}. All these architectures increasingly embed physically motivated priors such as long-range electrostatics or dispersion directly into the model, improving data efficiency and the reliability of out-of-distribution predictions~\cite{P6587}.

\begin{figure*}[!ht]
    \centering
    \includegraphics[width=\linewidth]{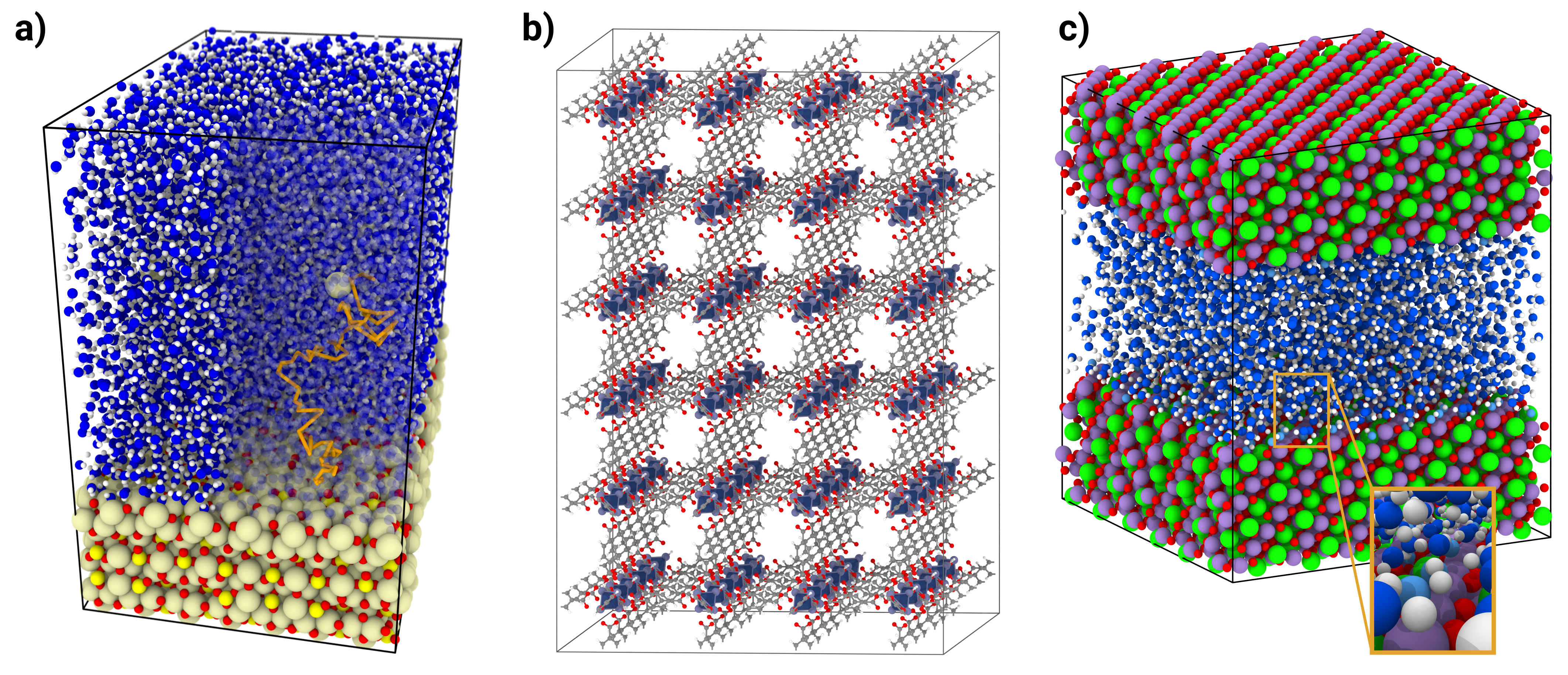}
    \caption{Illustrations of materials and system sizes \RuNNer has been designed for. \textbf{a)} Diffusion of a calcium ion at a calcium silicate-water
    interface (hydrogen: white, surface oxygen: red, water oxygen: blue, silicon: yellow, calcium: ochre)~\cite{prus2025moleculardynamicssimulationsgammabelite010water}. \textbf{b)} Metal organic framework MOF-5~\cite{MOF_structure,P5537} (hydrogen: white, carbon: grey, oxygen: red, zinc: purple. \textbf{c)} Water dissociation at the LiMn$_2$O$_4$--water interface~\cite{P6141} (hydrogen: white, lithium: purple, surface oxygen: red, water oxygen: blue, hydroxide oxygen: cyan, manganese: purple).}
    \label{fig:intro}
\end{figure*}

High-dimensional neural network potentials (HDNNPs), introduced in 2007~\cite{P1174}, have been the first MLP type applicable to large systems containing thousands of atoms, such as the systems exemplified in Fig.~\ref{fig:intro}. In the following years, HDNNPs have been continuously further developed to extend their scope and applicability. To enable a clear assessment of the physics included in the different HDNNP models, recently a classification scheme into different generations has been introduced~\cite{P6018}, which can also be applied to many other types of MLPs. 

The original HDNNPs introduced in 2007 are now classified as second-generation (2G) HDNNPs for a distinction from earlier first-generation neural network potentials that are restricted to low-dimensional systems.
2G-HDNNPs proposed the representation of the potential energy as a sum of environment-dependent atomic energy contributions, each predicted by a feed-forward neural network from atomic features describing the local atomic environments. Third-generation (3G) HDNNPs~\cite{P3132,P2962} augment this short-range energy by a long-range electrostatic term based on environment-dependent partial charges predicted by an additional set of atomic networks. Moreover, dispersion interactions beyond the local environments can be included.
However, 3G models still assume that charges are determined solely by the local chemical environment. Therefore, they are unable to describe non-local charge transfer, i.e., processes in which the charge distribution in the system changes in response to a far-away perturbation. 
Fourth-generation (4G) HDNNPs~\cite{P4419,P5932,P5977} overcome this restriction by allowing charge to redistribute according to the global structure and composition of the system using a global charge equilibration (QEq) scheme~\cite{P1448}. The ability to describe such global phenomena is essential for systems such as charged defects in ionic crystals, metal atoms adsorbed on doped oxide surfaces, or the protonation state of organic molecules.

While the HDNNP method has been developed in the RuNNer code -- the \emph{Ruhr University Neural Network energy representation} -- starting almost two decades ago, in the following years due to the popularity of the method, many other HDNNP codes have been published, such as AMP~\cite{P4551}, \AE{}net~\cite{P4637}, ANI-1~\cite{P4945}, Tensormol~\cite{P5313}, SIMPLE-NN~\cite{P5816}, n2p2~\cite{P5603}, PiNN~\cite{P5719}, PANNA~\cite{P6532}, and Fortnet~\cite{P6350}, and each of these software packages has its specific advantages and limitations. 
In spite of this variety of codes, recent methodological extensions like the development of 4G-HDNNPs require substantial structural changes in the software architecture, and consequently older codes like RuNNer 1.3 and most of the other HDNNP codes do not offer a high-performance implementation suitable for routine production calculations of all HDNNP generations on equal footing.

In this work, we present \RuNNer, the successor of RuNNer 1.3~\cite{P4444,P5128}. The code has been completely redesigned from scratch to optimize its performance taking recent advances with a particular focus on 4G-HDNNPs into account. We have chosen an implementation in modern Fortran (2003/2008 standards) for its maturity, high performance, support by multiple compilers and the long-term availability and stability of Fortran-based software.
A quasi-linear-scaling plane-wave QEq solver reduces the cost of global charge equilibration, a key step in 4G-HDNNPs, from $\mathcal{O}(N^3)$ to $\mathcal{O}(N\log^2 N)$. This algorithmic advance, combined with a hybrid MPI/OpenMP parallelization scheme and a memory management strategy that removes the training overhead historically associated with long-range interactions, makes large-scale 4G-HDNNP molecular dynamics and training computationally tractable for the first time. Committee models are natively supported across all training and prediction workflows via highly efficient implementations that maximize computation sharing. \RuNNer further provides general, extensible interfaces upon which specific integrations are built. This includes native support for LAMMPS and the Atomic Simulation Environment (ASE), exposing all of these capabilities, including committee-based uncertainty quantification, to two of the most widely used atomistic simulation frameworks, and is complemented by a Python workflow layer for dataset management and training analysis.

The paper, which describes this implementation, is structured as follows: Section~\ref{sec:theory} introduces the theoretical framework underlying all HDNNP generations implemented in \RuNNer, including the energy expressions, descriptor formalisms, long-range interaction models, and optimization objective. Section~\ref{sec:implementation} describes how this framework is translated into an efficient, modular software architecture. We then present extensive performance benchmarks in Section~\ref{sec:performance} before demonstrating the practical importance of the 4G framework for a specific example, an extension of the Au/MgO(Al$^*$) non-local charge transfer benchmark~\cite{P5932}.

\section{Methods}
\label{sec:theory}

\subsection{High-Dimensional Neural Network Potentials}

\begin{figure*}[p]
    \centering
    \includegraphics[width=\linewidth]{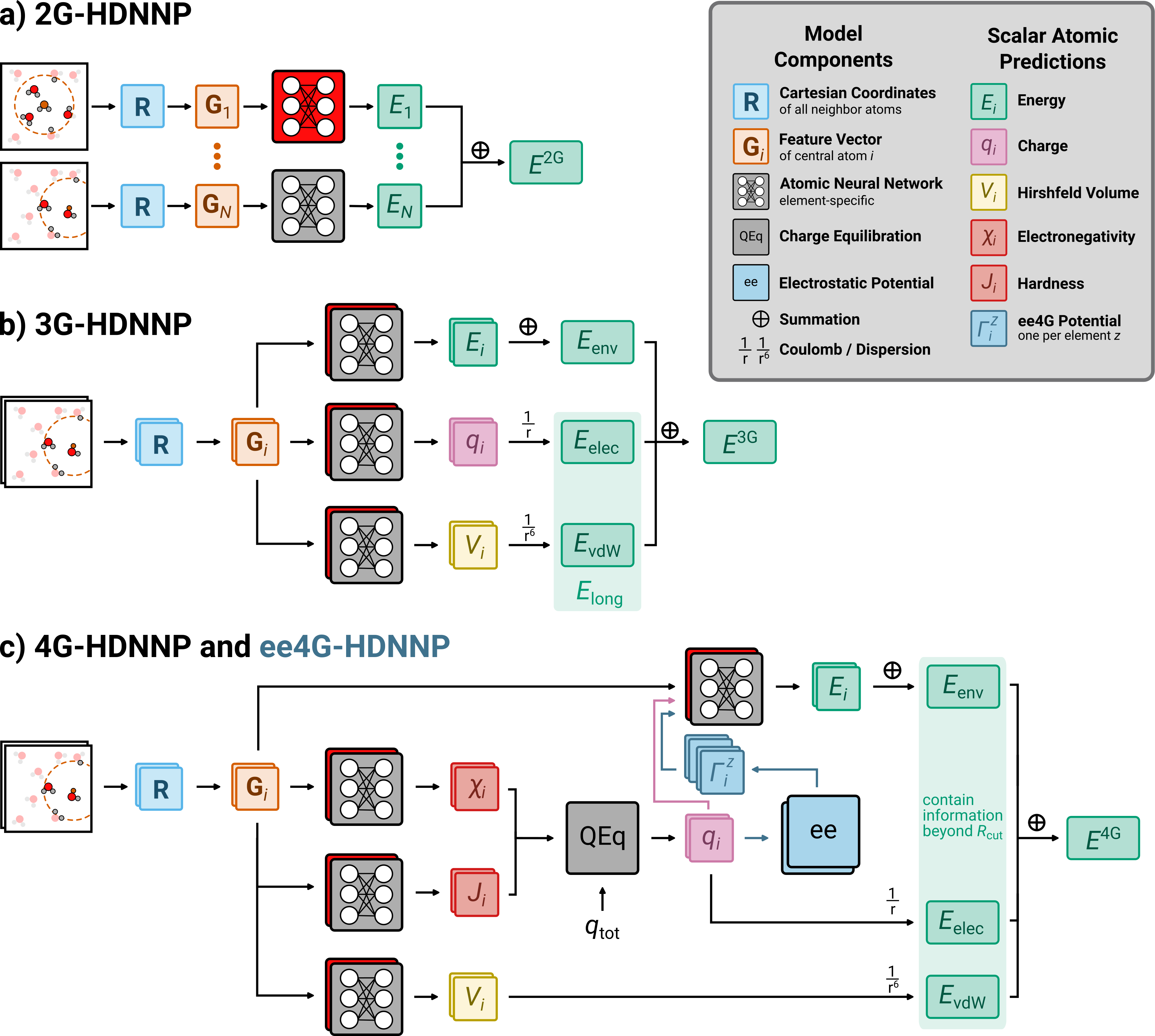}
    \caption{Flowcharts depicting the different generations of HDNNPs. \textbf{a)} In 2G-HDNNPs the local atomic environments of the central atoms $i$ (orange) are defined by a cutoff radius (dashed orange circle). The Cartesian coordinates of all neighbor atoms ($\mathbf{R}$) are used to construct feature vectors $\mathbf{G}_i$, which feed into element-specific atomic neural networks (here, e.g., red for oxygen, grey for hydrogen). Each network predicts the respective atomic energy $E_i$, and these energies are combined to yield the total energy $E^{\mathrm{2G}}$. \textbf{b)} In 3G-HDNNPs, feature vectors are constructed similarly (where we have compressed the information for different atoms into a stacked representation for brevity), but feed into several groups of element-specific neural networks to predict distinct atomic properties. The atomic charges $q_i$ determine the electrostatic energy $E_{\mathrm{elec}}$ via Coulomb interactions, while atomic Hirshfeld volumes $V_i$ serve as the basis for the van der Waals dispersion contribution $E_{\mathrm{vdW}}$.
    The atomic energies $E_i$ yield the local energy $E_{\mathrm{env}}$, which represents the part of the energy not covered by $E_{\mathrm{long}}$.
    \textbf{c)} 4G-HDNNPs use the atomic feature vectors to predict atomic electronegativities $\chi_i$ and hardnesses $J_i$. These parameters form the basis of a charge equilibration (QEq) scheme that distributes the total charge $q_{\mathrm{tot}}$ across all atoms. The resulting atomic charges $q_i$ encode global information about the charge distribution throughout the system. In a standard 4G model, this non-local information channel feeds directly into the atomic neural networks as additional input feature to predict consistent atomic energies $E_i$ forming the environment-dependent energy contribution $E_{\mathrm{env}}$. Combined with the electrostatic interaction of the equilibrated charges $E_{\mathrm{elec}}$ and the 3G-like vdW term $E_{\mathrm{vdW}}$, the total energy $E^{\mathrm{4G}}$ captures both short-range and long-range effects. Panel \textbf{c)} also illustrates the ee4G-HDNNP architecture, which extends the 4G model by introducing electrostatic embedding (ee) channels. Here, in addition to feeding the atomic charges directly into the atomic energy network, the $q_i$ are also used to compute element-specific electrostatic potentials ($\Gamma_i^z$) serving as additional non-local input features for the atomic energy predictions.}
    \label{fig:hdnnp}
\end{figure*}

An overview of second-generation (2G), third-generation (3G) and fourth-generation (4G) HDNNPs is provided in Fig.~\ref{fig:hdnnp}. A concise summary of these methods will be given in the next sections to introduce the essential background needed for the discussion of the implementation in \RuNNer. A more detailed description of HDNNPs, the underlying methods and the training procedures can be found in several comprehensive reviews~\cite{P6018,P5128,P4444,P6548,P4106}.

\subsubsection{2G-HDNNPs}

2G-HDNNPs are local MLPs based on the assumption that the majority of atomic interactions is short-ranged. The total potential energy $E^{\mathrm{2G}}$ of 2G-HDNNPs for a system containing $N$ atoms is given by the sum of the atomic energy contributions $E_i$~\cite{P1174},
\begin{equation}
E^{\mathrm{2G}} = \sum_{i=1}^{N} E_i\left(\mathbf{G}_i\left(\mathbf{R}\right)\right)\,,
\end{equation}
where each $E_i$ is predicted by an element-specific feed-forward neural network as a function of a feature vector $\mathbf{G}_i$. This vector encodes the local chemical environment within a cutoff radius $r_{\mathrm{cut}}$ based on the atomic coordinates of the neighbor atoms $\mathbf{R}=\{\mathbf{r}_i\}$. The features are constructed to fulfill the mandatory invariances of the potential energy with respect to translation, rotation, and permutation~\cite{P2882}. There is one neural network evaluation per atom in the system (Fig.~\ref{fig:hdnnp}a) such that the computational cost scales linearly with system size. Analytical forces $\mathbf{F}_j$ are obtained by chain-rule differentiation through the feature maps,
\begin{equation}
\mathbf{F}_j = -\nabla_{\mathbf{R}_j} E^{\mathrm{2G}}
= -\sum_{i=1}^N \sum_{\mu=1}^{M_i} \frac{\partial E_i}{\partial G_{i, \mu}} \frac{\partial G_{i, \mu}}{\partial \mathbf{R}_j} \,,\label{eq:2G_forces}
\end{equation}
where $G_{i, \mu}$ denotes the $\mu$-th component of the $M_i$-dimensional atomic feature vector $\mathbf{G}_i$. Consequently, vectorial or tensorial properties, e.g., the stress tensor, derived from the atomic energies preserve equivariance.
The parameters of the neural networks are adjusted in an iterative training procedure such that the sum of the learned atomic energies matches the total energies of a set of reference electronic structure calculations. In addition, atomic forces are typically used for training the neural network weight parameters.

\subsubsection{3G-HDNNPs}

Third-generation (3G) HDNNPs extend the energy expression by incorporating explicit long-range interactions beyond the cutoff radius, which are included in 2G-HDNNPs only implicitly in an averaged way. A defining characteristic of 3G-HDNNPs is that the long-range energy $E_{\mathrm{long}}$ is expressed through interactions based on environment-dependent atomic properties $\mathbf{P}=\{P_i\}$ that are predicted by additional atomic neural networks as local quantities (Fig.~\ref{fig:hdnnp}b),
\begin{equation}
E^{\mathrm{3G}} = \sum_{i=1}^{N} E_i\left(\mathbf{G}_i\right) + E_\mathrm{long}\left(\mathbf{P}(\mathbf{G})\right) \,,
\label{eq:3g}
\end{equation}
where we have dropped the dependence of the feature maps $\mathbf{G}$ of all atoms on the coordinates $\mathbf{R}$ for better readability.
While the original 3G-HDNNP formulation focused on electrostatic interactions mediated by environment-dependent atomic charges~\cite{P2962,P3132}, \RuNNer generalizes this concept to arbitrary long-range interaction terms, since also the tails of dispersion interactions beyond the cutoff can become important in some cases. The energy of 3G-HDNNPs is given by
\begin{equation}
E^{\mathrm{3G}} = \sum_{i=1}^{N} E_i\left(\mathbf{G}_i\right)
+ E_{\mathrm{elec}}\left(\mathbf{q}\left(\mathbf{G}\right)\right)
+ E_{\mathrm{vdW}}\left(\mathbf{V}\left(\mathbf{G}\right)\right)\,,
\end{equation}
where $\mathbf{q}$ and $\mathbf{V}$ denote the machine-learned atomic charges for electrostatics and Hirshfeld volumes for van der Waals energies~\cite{PhysRevB.104.054106, P2121} of all atoms, respectively. Therefore, while the machine-learned properties $\mathbf{P}$ are strictly local, the resulting long-range interactions can have an infinite range, e.g., when using Ewald summation for electrostatics, or a much increased range in case of dispersion. In case of electrostatics, an optional total charge constraint may be applied to the learned atomic charges during inference.

Training of 3G-HDNNPs is commonly performed in a sequential manner. First, the auxiliary property models expressing $\mathbf{P}$ are trained against reference quantities obtained from electronic-structure calculations. The resulting long-range contributions are then evaluated and subtracted from the total reference energies and forces, allowing the short-range model to be trained to the residuals. This decomposition improves learning efficiency and avoids double counting of long-range interactions.

\subsubsection{4G-HDNNPs}

In spite of including long-range interactions, 3G models are still ``local'' since even interactions beyond the cutoff exclusively rely on environment-dependent atomic properties. Therefore, in general 2G and 3G MLPs are not applicable to systems with non-local dependencies like long-range charge transfer~\cite{P5977}. 4G-HDNNP~\cite{P5932,P5977} models overcome this limitation by introducing globally equilibrated charges~\cite{P1448,P4419}.
The charge equilibration (QEq)~\cite{P1448} step in 4G-HDNNPs employs atomic electronegativities $\chi_i$ and hardnesses $J_i$, which are learned in the training process (Fig.~\ref{fig:hdnnp}c). QEq then distributes the total charge of the system $q_{\mathrm{tot}}$ among the atoms such that the charge distribution minimizes an electrostatic energy functional. This procedure is described in detail in Sec.\;\ref{sec:electrostatics}.
The resulting atomic charges $q_i$ are used for two purposes: the calculation of the long-range electrostatic energy and to extend the atomic feature vectors by including global information to overcome the strict locality of the atomic energies $E_i$. 
Moreover, for non-periodic systems the QEq step, which adds information about the total charge $q_{\mathrm{tot}}$ of the system, generalizes the applicability of HDNNPs to multiple total charge states.

The energy $E^{\mathrm{4G}}$ is defined as
\begin{align}
E^{\mathrm{4G}} = & \sum_{i=1}^{N} E_i\left(\mathbf{G}_i,\,q_i\left(\mathbf{G},q_{\mathrm{tot}}\right)\right) \\\notag
    & + E_{\mathrm{elec}}\left(\mathbf{q}\left(\mathbf{G},q_{\mathrm{tot}}\right)\right)
+ E_{\mathrm{vdW}}\left(\mathbf{V}\left(\mathbf{G}\right)\right)\,.
\end{align}

The \emph{electrostatically embedded} variant (ee4G-HDNNP~\cite{P6542}) additionally appends a vector of element-specific electrostatic potentials at each atomic site to the feature vector, providing additional information about the local electrostatic environment resulting from the equilibrated charges.
Like 3G-HDNNPs, fourth generation models are typically trained in two steps.

\subsection{Atomic Features}
\label{sec:theory_features}

The feature vector $\mathbf{G}_i$ encodes the local environment of atom $i$ using features that are invariant under translation, rotation, and permutation of chemically equivalent neighbors within the cutoff radius. \RuNNer provides two classes of features: atom-centered symmetry functions and overlap matrix fingerprints.

\subsubsection{Atom-Centered Symmetry Functions}
\label{sec:theory_acsf}

The atom-centered symmetry function (ACSF)~\cite{P2882} features implemented in \RuNNer comprise radial and angular functions encoding the positions of the neighboring atoms up to the cutoff radius, which is typically chosen between 5 and 10~\AA{}. 

Radial ACSFs probe the element-resolved radial neighbor densities. The most commonly used radial function has the form (labeled type 2 following the notation in ~\citenum{P2882})
\begin{equation}
G_{i,\mu}^{(2)} = \sum_{j \in \mathcal{N}_i} \exp\!\bigl(-\eta_\mu(r_{ij}-R_{\text{shift},\mu})^2\bigr)\, f_\mathrm{cut}(r_{ij}) \,,
\label{eq:g2}
\end{equation}
where $\eta_{\mu}$ controls the Gaussian width, $R_{\text{shift},\mu}$ positions the center, $f_\mathrm{cut}$ is a smooth cutoff function, $j$ runs over all neighbors $\mathcal{N}_i$ and $\mu$ separates functions with different parameters. Additionally, \RuNNer provides a simple counting function, which is the element-specific sum of the plain cutoff functions with respect to all neighbors of a given element. 
Angular ACSFs encode the angular distribution of the neighboring atoms and three types are available: type 3, type 8, and type 9. The functional forms of all ACSFs are provided in the SI, Sec.~\ref{section:si:features:acsf}, and a detailed description of their properties can be found in Ref.~\citenum{P2882}. 

In their original formulation, different chemical elements are distinguished by constructing separate, element-wise sets of ACSFs, each evaluated with its own set of parameters such as $\eta_\mu$ and $R_{\mathrm{shift},\mu}$. Alternatively, all ACSFs in \RuNNer can be augmented by element-embracing prefactors~\cite{eckhoff_lifelong_2023} that directly encode the chemical identity of the neighboring atoms, avoiding the need for using element-wise feature sets. For magnetic systems, spin-dependent prefactors modulate each pairwise or triplet contribution according to the local spin embedding~\cite{P6057}.

\subsubsection{Overlap Matrix Fingerprints}
\label{sec:theory_om}

\RuNNer provides an implementation of atom-centered overlap matrix (OM) fingerprints, which are well-established features for environment comparison and structural discrimination~\cite{sadeghiMetricsMeasuringDistances2013,parsaeifardAssessmentStructuralResolution2021} and have been used in the context of MLPs, e.g., in the graph neural network EOSnet~\cite{P7370}.

In \RuNNer, the OM is evaluated for each atomic environment as a \emph{feature map}, which is differentiable with respect to atomic positions to enable consistent force predictions. 
Since all neighbors in the local environment contribute to the atomic OM employing Gaussians of element-specific widths, the OM representation avoids the explicit combinatorial bookkeeping required by conventional element-specific features.
Specifically, at each neighbor site $j$ we place a minimal set of $n_{\mathrm{orb}}$ normalized atom-centered Gaussian-type functions $\Gamma_{j\nu}(\mathbf{r})$, where $\nu$ enumerates the members of the local GTO basis, i.e., basis functions distinguished by their angular character and Gaussian width.
The overlap matrix element for the environment of atom $i$ referring to atom pair $jk$, where $j$ and $k$ include all neighbors and the central atom $i$, is then given by
\begin{equation}
S^{\,i}_{j\nu,k\mu} \;=\; \int \Gamma_{j\nu}(\mathbf{r})\, \Gamma_{k\mu}(\mathbf{r})\, \mathrm d\mathbf{r}\,,
\label{eq:om_overlap_runner}
\end{equation}
yielding a symmetric matrix $\mathbf{S}^{i}\in\mathbb R^{n^i_\mathrm{bas}\times n^i_\mathrm{bas}}$, with $n^i_\mathrm{bas}=(|\mathcal{N}_i|+1)n_{\mathrm{orb}}$.
While this matrix is already restricted to the local environment through the neighbor set $\mathcal N_i$, a simple truncation of the environment at a finite cutoff would cause the matrix to change discontinuously when an atom enters or leaves the environment.
Each atom-centered basis function is hence weighted by the cutoff factor associated with its distance from the central atom yielding

\begin{equation}
\widetilde S^{\,i}_{j\nu,k\mu}
\;=\;
f_{\mathrm{cut}}(r_{ij}) \; S^{\,i}_{j\nu,k\mu}\; f_{\mathrm{cut}}(r_{ik}) \,.
\label{eq:om_cutoff_runner}
\end{equation}
The OM features are then constructed from the eigenvalue spectrum $\mathbf{\uplambda}^i$ of $\mathbf{\widetilde S}^{\,i}$,
\begin{equation}
\mathbf{\Lambda}^{\,i} = \left(\mathbf{U}^{\,i}\right)^T \mathbf{\widetilde S}^{\,i}\mathbf{U}^{\,i}\,, \qquad
\mathbf{\uplambda}^{\,i} =\mathrm{diag}(\mathbf{\Lambda}^{\,i})\,,
\label{eq:om_eigs_runner}
\end{equation}
where $\mathbf{\Lambda}$ is the diagonal matrix of eigenvalues and the columns of $\mathbf{U}$ are orthonormal eigenvectors. The first $n_{\lambda}\leq n^i_{\mathrm{bas}}$ entries of $\mathbf{\uplambda}^i$ then yield a sorted and truncated set of eigenvalues as the feature vector $\mathbf{g}^i$ of atom $i$.
In the context of MLPs, however, this representation still has a critical drawback:
since eigenvalues may cross for some structural changes, discontinuities may be introduced in the first feature derivatives, which is problematic for force learning.

To make the OM a robust feature vector for MLPs, \RuNNer employs a \emph{smooth spectral projection} that eliminates the eigenvalue-crossing by replacing the ordered list of eigenvalues by an order-invariant, smooth function of the spectrum.
We define a fixed grid $\{x_{l^{\prime}}\}_{{l^{\prime}}=1}^{n_g}$ of $n_g$ points in $[0,L]$, e.g., $x_{l^{\prime}} = L\,{l^{\prime}}/n_g$, and map the spectrum to features via

\begin{equation}
\widetilde{g}^{\,i}_{l^{\prime}} \;=\; \sum_{l=1}^{n_\lambda} \sin\!\bigl(\lambda^{\,i}_l \, x_{l^{\prime}} \bigr)\,,
\qquad l^{\prime} =1,\dots,n_g \,,
\label{eq:om_gridmap_runner}
\end{equation}
This transformation has two main consequences for MLPs:
(i) the obtained feature vector is invariant under permutations of eigenvalues (hence no sorting is required), and
(ii) features are smooth in $\lambda^i$ even at eigenvalue crossings of the original OM.
Consequently, the feature derivatives are well-defined and continuous. The calculation of derivatives is presented in the SI, Sec.~\ref{section:si:features:overlap_matrix:derivatives}.

\subsection{Electrostatics and Charge Equilibration}
\label{sec:electrostatics}

Both, 3G- and 4G-HDNNPs, require an efficient evaluation of long-range Coulomb interactions. \RuNNer expresses the electrostatic energy 
using a Coulomb kernel $\mathbf{E}$ based on Gaussian charge distributions of element-specific width $\sigma$,
\begin{equation}
    E_\text{elec} = \frac{1}{2}\mathbf{q}^{\mathrm{T}}\mathbf{E}\mathbf{q}\,,
    \label{eq:elec_energy}
\end{equation}
with the matrix elements $E_{ij}$ between two normalized Gaussians defined as
\begin{equation}
E_{ij} = 
\begin{cases}
    \dfrac{1}{\sigma_i \sqrt{\pi}}\,, & \text{if } i = j \\[10pt]
    \dfrac{\text{erf}\left( \dfrac{r_{ij}}{\sqrt{2}\gamma_{ij}} \right)}{r_{ij}}\,, & \text{otherwise}
\end{cases}
\end{equation}
and $\gamma_{ij} = \sqrt{\sigma_i^2+\sigma_j^2}$.


The Gaussian smearing preserves the correct $1/r$ behavior at large separations while removing the short-range divergence. Additionally, a screening function can be applied inside the atomic environments to dampen the electrostatic energy at distances that can be fully described by the atomic energies.

In third-generation models, the atomic charges $q_i$ are learned as local environment-dependent properties subject to an overall charge constraint. 4G-HDNNPs introduce a charge equilibration step based on the QEq scheme by Rappe \textit{et al.}~\cite{P1448}, which distributes the total charge of the system by minimizing the energy expression
\begin{equation}
    E_\text{QEq} = \frac{1}{2}\mathbf{q}^{\mathrm{T}}\mathbf{A}\mathbf{q} + \mathbf{q}^{\mathrm{T}}\boldsymbol{\upchi}\,,
\label{eq:qeq_energy}
\end{equation}
with
\begin{equation}
\mathbf{A} = \mathbf{E} + \mathbf{J}^{\mathrm{T}}\mathbf{I}\,.
\end{equation}
In this second order Taylor expansion of the total energy the first order expansion coefficient is the vector of atomic electronegativities $\boldsymbol{\upchi}$, while the second order prefactor is the vector of atomic hardnesses $\mathbf{J}$. $\mathbf{I}$ is the identity matrix.

Setting $\frac{dE_{\mathrm{QEq}}}{d\mathbf{q}} = 0$ leads to the system of linear equations
\begin{equation}
\left(
\begin{array}{c|c}
\mathbf{A} & \begin{matrix} 1 \\ \vdots \\ 1 \end{matrix} \\
\hline
\begin{matrix} 1 & \dots & 1 \end{matrix} & 0
\end{array}
\right)
\left(
\begin{array}{c}
q_1 \\
\vdots \\
q_N \\
\hline
\lambda
\end{array}
\right)
=
\left(
\begin{array}{c}
-\chi_1 \\
\vdots \\
-\chi_N \\
\hline
q_{\text{tot}}
\end{array}
\right)\,,
\end{equation}
or
\begin{equation}
\mathbf{A'}\mathbf{q'} = -\boldsymbol{\upchi'}\,,
\label{eq:4G_qeq_prime}
\end{equation}
where the Lagrange multiplier $\lambda$ is chosen such that the charge constraint $\sum_i^N q_i = q_{\mathrm{tot}}$ is fulfilled.

To ensure that $\mathbf{A}$ stays positive definite, $\mathbf{J} > 0$ must be enforced. In \RuNNer, this can be done by applying activation functions to the output of the hardness model. Electronegativities are typically not constrained and may become negative, which is physically reasonable for electropositive species~\cite{P4419}. We find that initializing $\chi$ and $\mathbf{J}$ with with values scaled between one and twenty times standard Pearson electronegativity and hardness values published in the literature yields optimal convergence toward physically meaningful parameters.

Calculating the derivatives of Eq.~\ref{eq:qeq_energy} to obtain the atomic force contributions formally requires the derivatives of the charges with respect to the atomic coordinates. For large-scale production calculations, this is prohibitively expensive since it requires solving $3N$ linear systems of equations. However, as first shown by Handy and Schaefer~\cite{handy_schaefer} and later transferred to the problem of charge equilibration by Poier \textit{et al.}~\cite{poier_lagrange}, we can use the Lagrange method of undetermined multipliers to obtain the forces by solving only a single system instead. This ``force trick'' is explained in more detail in the SI, Sec.~\ref{section:si:elec:4G}.
\subsection{van der Waals Interactions}
\label{sec:vdw}

Dispersion interactions are incorporated in \RuNNer through a Hirshfeld volume-based scheme~\cite{PhysRevB.104.054106} in which environment-dependent Hirshfeld volumes $V_i$ are predicted by atomic neural networks. Following the Tkatchenko-Scheffler method~\cite{P2121}, these volumes can be employed to scale the free-atom $C_6$ coefficients to obtain pair-wise dispersion energies adjusting to the local chemical environments,
\begin{equation}
  C_6^{ij} = C_6^{ij,\mathrm{free}} \cdot V_i \cdot V_j \,.
\end{equation}
The dispersion energy is then the sum of the scaled pairwise $C_6/r^6$ contributions over all atom pairs with appropriate damping to avoid short-range divergence~\cite{P7126}. Dispersion interactions are typically truncated beyond a very large cutoff radius that must be converged for the system of interest.
Alternatively, in particular if DFT reference data is used, HDNNP energies can be combined with established dispersion corrections like the D3 method~\cite{P3112}.

\subsection{Two-Body Repulsion}

\RuNNer provides an implementation of the Ziegler--Biersack--Littmark (ZBL) potential~\cite{ziegler_stopping_1985} (see SI, Sec.~\ref{section:si:twobody:zbl}, for details). This enables adding short-range nuclear repulsion between atoms to any HDNNP generation. 

\subsection{Parameter Optimization}
\label{sec:theory_optimization}

Neural network parameters are adjusted by minimizing a loss function that compares predicted energies, forces, and atomic properties with reference electronic-structure data. \RuNNer supports gradient descent (GD), stochastic gradient descent (SGD), Adam~\cite{P5404}, and two variants of the extended Kalman filter (standard and fading-memory)~\cite{P1308}. The mathematical details of each optimizer, including the Kalman filter update equations and hyperparameter scheduling, are provided in the SI, Secs.~\ref{section:si:training:loss} and \ref{section:si:training:opt}.

\section{Implementation Details}
\label{sec:implementation}

In this section we describe how the theoretical framework of Section~\ref{sec:theory} is efficiently translated into the \RuNNer software. We cover the overall code architecture, parallelization and memory strategies, the specific model and feature types available, the electrostatics solver choices, and the interfaces to external simulation software.

\subsection{Four-Level Architecture}

\begin{figure*}[!ht]
    \centering
    \includegraphics[width=\linewidth]{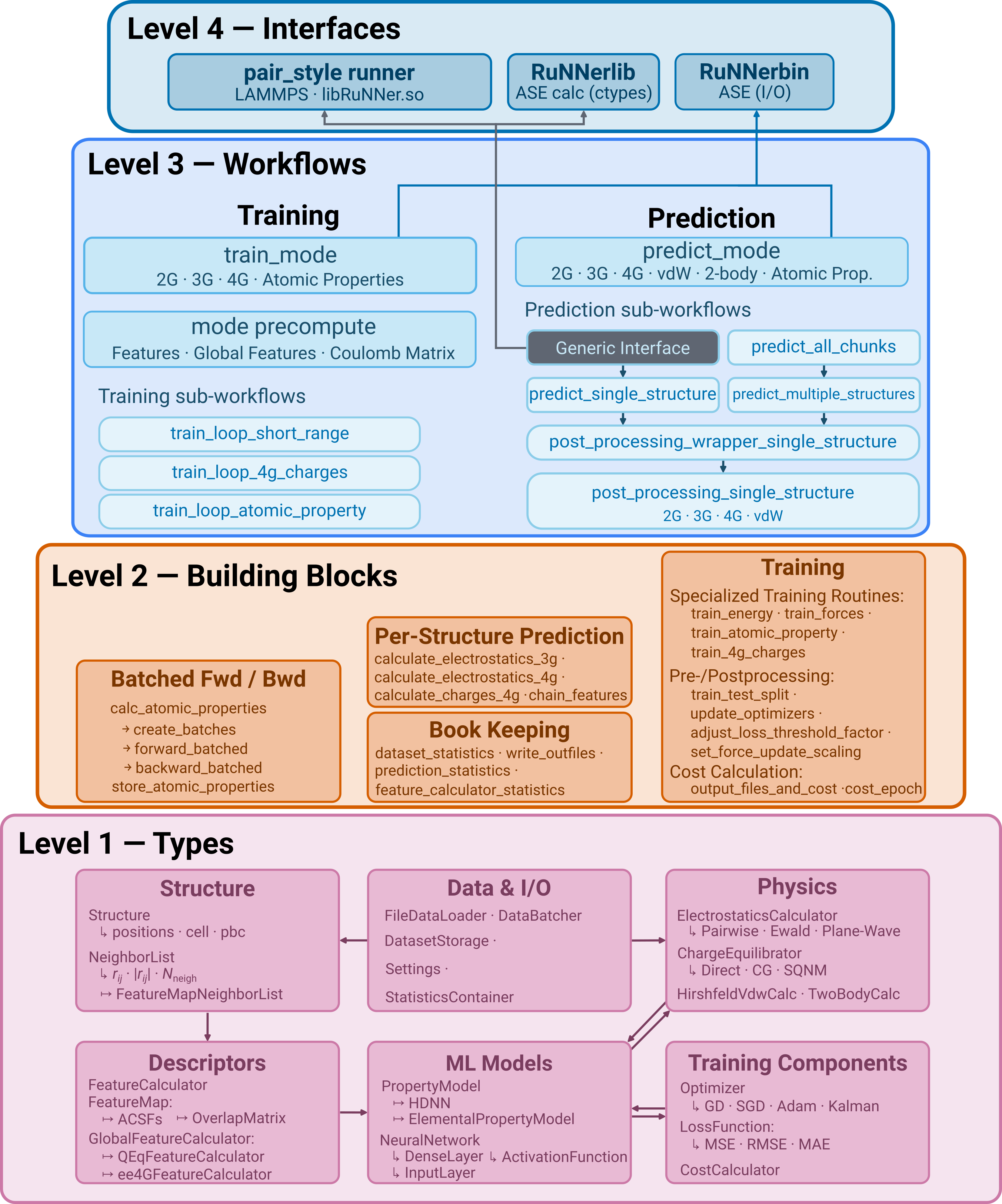}
    \caption{Overview of the \RuNNer suite and its components organized in a hierarchy of four levels. Self-contained, object-oriented types (\textbf{Level 1}) form the basis of the program, whose state is altered through independent, functional building blocks (\textbf{Level 2}). Complex tasks are managed through procedural workflows (\textbf{Level 3}), which also form the capabilities of the standalone \RuNNer binary. The individual workflows are exposed through multiple C-bound interfaces (\textbf{Level 4}), that make them easily accessible by other languages and tools. We provide custom embeddings for Python, the Atomic Simulation Environment (ASE), and LAMMPS. A detailed explanation of each level is given in the main text.}
    \label{fig:flowchart}
\end{figure*}

The central design challenge of \RuNNer has been to reconcile two very different requirements: the high-performance implementation needed for large-scale atomistic simulations and the flexibility needed to rapidly prototype new methodical developments. The solution is a hierarchical structure consisting of four levels. 

\subsubsection{Level 1}

At the lowest level, self-contained object-oriented types encapsulate individual data structures and their immediate operations (Fig.~\ref{fig:flowchart} \--- Level 1). For example, \RuNNer provides a generally applicable neural network type with attached weights and biases, a feature calculator managing feature map evaluation, and a data loader with automatic memory budget determination. These types are designed to be as small and independent as possible. They communicate through simple, intrinsic data types rather than shared mutable states, which enables easy extensions. We favor composition over inheritance: advanced capabilities are built by embedding simple types within larger ones, rather than by constructing deep class hierarchies. This way, the individual components stay light-weight and modular, making them easy to understand, to extend, and to replace.

\subsubsection{Level 2}

The second level consists of building blocks that combine multiple types to perform mathematically intensive operations, such as batched forward and backward passes, chain rule applications, parameter updates, or dispersion energy calculations (Fig.~\ref{fig:flowchart} \--- Level 2). Each routine in this layer is self-contained, making the building blocks reusable across different workflows and simplifying incremental code changes like performance enhancements, future GPU acceleration, or new methodological developments.

\subsubsection{Level 3}

The third level comprises workflows, which are procedural routines that initialize types, schedule calls to building blocks, and handle input and output (Fig.~\ref{fig:flowchart} \--- Level 3). These workflows are the public-facing API. Since the underlying building blocks are stateless and context-free, adding a new physical term requires only a new building block and its insertion into the relevant workflow, with no structural coupling to any other part of the code. \RuNNer can therefore combine 2G short-range energies, electrostatics (3G or 4G), Hirshfeld volume-based dispersion corrections, and short-range repulsion within a single prediction pipeline: each contribution is independently developed and tested at the type or building-block level, and the combination is controlled by workflow logic.

\subsubsection{Level 4}

The three levels discussed so far form the \RuNNer library, which is the basis of the standalone executable program. The highest level four contains the interfaces through which users may interact with the library or the binary application (Fig.~\ref{fig:flowchart} \--- Level 4). We provide a native pair style for LAMMPS (s. Section~\ref{sec:lammps}) and a generic Python interface that is fully integrated into the atomic simulation environment (ASE)~\cite{P5915} as a \textsc{Calculator} object (s. Section~\ref{sec:ase_interface}). Both access \RuNNer through a shared library object. Furthermore, \RuNNer can be accessed through Python in the form of the ASE-like I/O calculator \textsc{RuNNerbin}. While the high-level interfaces are usually specific to the calling code, they all access the same library routines. This way, all HDNNP generations are natively supported in all interfacing codes. The interfaces are described in more detail in Sec.\;\ref{sec:interfaces}.

\subsection{Parallelization Strategy and Batching}

\begin{figure*}[!ht]
    \centering
    \includegraphics[width=\linewidth]{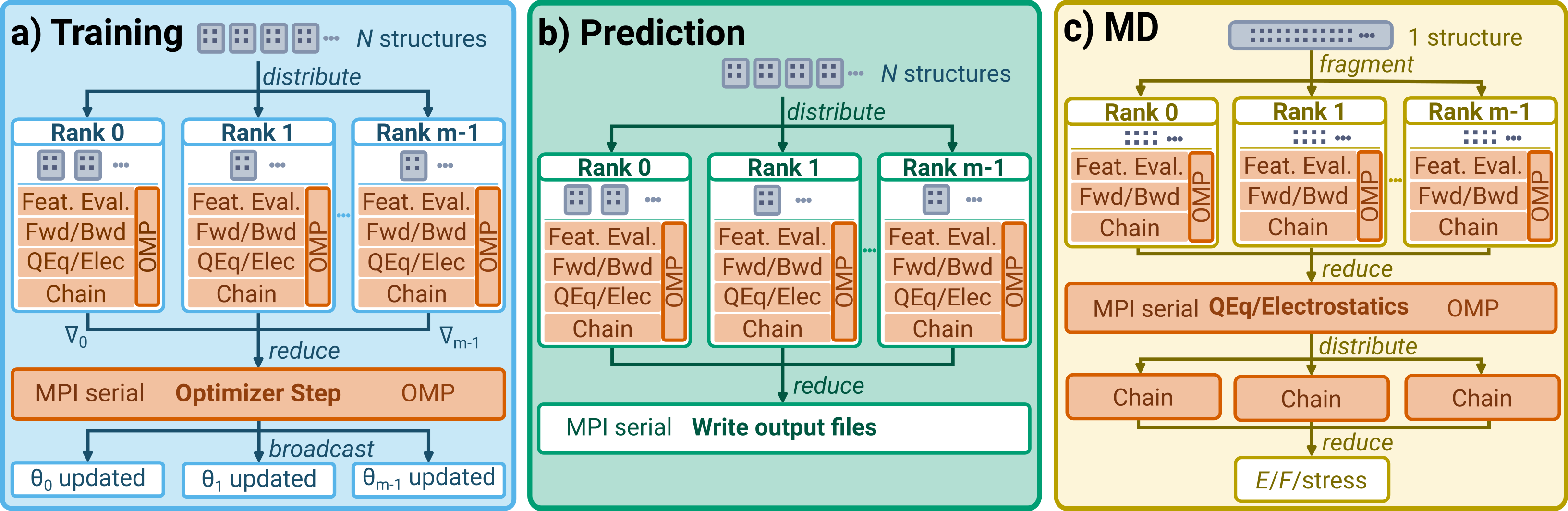}
    \caption{Hybrid MPI/OpenMP parallelization strategy in \RuNNer for the three main workflows training, prediction, and molecular dynamics (MD). During training (left panel), structures are distributed across MPI tasks, where each task handles the per-batch operations like feature evaluation, forward/backward pass, charge equilibration (QEq), electrostatic evaluation, and chain-rule application. All of these operations are parallelized across threads via OpenMP. The calculated gradients $\theta$ are collected on the root MPI task and a collective optimizer step is performed before the updated parameters are redistributed to the MPI tasks and the next batch is loaded. When predicting one chunk of a dataset using the standalone \RuNNer binary (middle panel), operations follow a similar scheme. After each task has calculated the requested properties (e.g., energies, forces, stress) for its subset of structures, the information is collected on the root MPI task and written to output files. When interfacing to \RuNNer through a molecular dynamics (MD) code like LAMMPS (right panel), usually only a single, large structure is predicted that is fragmented across MPI tasks. Each task uses OpenMP to compute features, perform the forward/backward passes through the models, and compute force and stress contributions by applying the appropriate gradient chain rules. For 3G and 4G electrostatics/charge equilibration, all local properties and their gradient contributions must be collected on the master MPI rank where OpenMP parallelization is performed before the results are redistributed and required gradient contributions can be added on each task. The resulting arrays are then reduced by the MD propagator to yield the final energies $E$, forces $F$, and lattice stress.}
    \label{fig:parallelization}
\end{figure*}

\RuNNer employs a two-level parallelization strategy: MPI (Message Passing Interface) distributes work across compute nodes (or processes within a node) and OpenMP exploits thread-level parallelism within each process.

\subsubsection{Training}

At startup, the master process reads the model parameters and dataset and broadcasts this information to all workers (Fig.~\ref{fig:parallelization} \--- Training). Structures are distributed across MPI ranks by round-robin assignment, which achieves approximate load balance when structures have similar atom counts and densities. Within each epoch, each MPI task processes its local subset: neighbor lists are generated serially for its structures, whereas feature evaluation over atoms and unrolled matrix multiplications for chain rules are OpenMP parallelized, and the linear-algebra kernels in the forward and backward passes are offloaded to a threaded BLAS/LAPACK library. Appropriate batching strategies improve performance by increasing the size of the parallel regions within a batch. After each optimizer step the gradients from all ranks are accumulated by a collective reduction, the master updates the network weights, and the updated parameters are broadcast before the next batch.

For Kalman filter optimizers the multistream variant of Singraber \textit{et al.}~\cite{singraber_parallel_2019} is used: each MPI rank provides one data stream and a single Kalman update step is performed, avoiding a sequential gradient loop. A similar procedure is applied to all other optimizers implemented in \RuNNer.

\subsubsection{Prediction}

When predicting a full dataset (Fig.~\ref{fig:parallelization} \--- Prediction), model weights and feature scaling data are broadcast to all ranks and the structure set is divided into memory-adapted chunks (see Section~\ref{sec:memory}) that are distributed among ranks. Similarly to training, all ranks build neighbor lists, evaluate features, perform the forward/backward pass, run charge equilibration and electrostatics calculations, and compute gradients in parallel under OpenMP. Forces are accumulated locally and no inter-rank communication is needed after the forward pass, up until the final results are collected by the master process and written to the output file.

While parallelization at the structure level among ranks is possible when predicting an entire dataset, this is not viable for molecular dynamics. When interfacing \RuNNer with a simulation propagator like LAMMPS (Fig.~\ref{fig:parallelization} \--- MD), typically a domain decomposition scheme is used where atoms and their neighbor lists are distributed across MPI tasks. In this case, each task handles the computation of feature values, model predictions, and initial gradient concatenation. For 3G and 4G models, the electrostatic and QEq steps require global knowledge of the simulation cell. In the LAMMPS interface this is handled through LAMMPS' native domain decomposition and ghost-atom communication layers, which assemble the full local-plus-ghost atom list on the root MPI task where we perform a single, global charge equilibration and electrostatics calculation, before distributing the resulting per-atom quantities back via LAMMPS' forward/reverse communication mechanism and continuing with the remaining calculation steps in parallel.

\subsection{Memory Management}
\label{sec:memory}

Training and prediction require holding neighbor lists, feature vectors, and their derivatives in memory simultaneously, which for large datasets can exceed the available RAM. \RuNNer addresses this challenge through a chunked data-loading strategy implemented in \texttt{FileDataLoader} (Fig.~\ref{fig:flowchart} \--- Level 1), which enables an optimal tradeoff between memory consumption and computational efficiency. This strategy ensures that \RuNNer scales across a wide range of hardware architectures.

\subsubsection{Training}

Before the training loop begins, \RuNNer estimates the RAM required by the six major components: the raw structure data, the neighbor lists, the precomputed feature vectors, their nuclear coordinate derivatives, and, for 4G models, the global charge features and the associated Coulomb matrix $\mathbf{A}$. The memory budget can be set explicitly or left to be queried automatically from the operating system. The dataset (structures and their neighbor lists) is partitioned into the smallest number of equal-sized chunks that fit within the budget.

When the entire dataset fits into a single chunk, the code dynamically attempts to precompute and cache all features and their derivatives before the loop over training epochs begins. Feature evaluation is then absent from the inner loop, saving computational time by avoiding to recalculate the same values every epoch and maximizing the benefit of the batched matrix operations and OpenMP parallelism. When multiple chunks are required, features are computed on-the-fly for each batch and released immediately after the optimizer step. The two approaches can be mixed at a fine-grained level: for example, feature vectors can be cached while their derivatives, which require substantially more memory, are recomputed on demand.

Storage space of environment-dependent features and their derivatives scales linearly with the number of atoms in the training dataset.
For 4G and ee4G force training, the additional features with a global dependency (charges, electrostatic embeddings) and their derivatives are equally cached. It is noted that in 4G models storage of the dense matrix $\frac{\partial \mathbf{q}}{\partial \mathbf{r}}$ scales $\mathcal{O}\left(N^2\right)$ with the number of atoms in the system. For 4G charge training, the Coulomb matrix and its inverse are precomputed and stored, and the required memory also scales quadratically with the number of atoms per structure (see Sec.\;\ref{performance:4G_charge_training} for a more comprehensive explanation).

\subsubsection{Prediction}

\RuNNer's prediction mode enables the efficient recalculation of entire datasets. This is often practically useful, e.g., for evaluating the performance of a trained potential on validation datasets.
In prediction mode the number of chunks is adjusted such that all feature values and their derivatives fit into memory for a single chunk. This maximizes the throughput for OpenMP-parallelizable operations like feature evaluation and neighbor list construction, which are performed for the entire chunk at once and then consumed batch by batch during the forward pass. Because structures are independent, each chunk can be distributed over MPI ranks without cross-structure communication. 

In simulation runs, i.e., typically for a single large system, each MPI task of \RuNNer manages memory as required for the part of the simulation box it owns. At the first simulation step, the code creates a buffered neighbor list and a static feature storage space. If the number of atoms or neighbors owned by the task changes during the simulation, these structures are resized with a preallocated safety margin. Because such reallocations are rare, this strategy yields a substantial performance increase compared to dynamic reallocation.

\subsection{Available Models}
\label{sec:impl_models}

\RuNNer adopts a modular model architecture that separates generic machine-learning models from representations of physical observables. Generic models, such as neural networks, provide reusable functionality for tasks including parameter management, forward evaluation, and gradient propagation. Property models build upon these components to represent specific quantities of interest, such as energies, forces, charges, or other atomic quantities.

To facilitate interoperability across workflows, all property models implement a common interface that defines the routines required for training and inference. This abstraction allows model components to be exchanged without modifications to the surrounding infrastructure and ensures consistent integration into optimization, prediction, and analysis workflows.

A prominent example is the implementation of \RuNNer's \textsc{HDNN} model type, in which species-specific neural networks are combined into a unified property model. In addition, \RuNNer provides generic elemental property models (\textsc{ElementalPropertyModel}) for fitting environment-independent quantities.

\subsubsection{Second-Generation HDNNPs}
\label{sec:features_2g}

In \RuNNer the second generation model is represented by a single \textsc{HDNN} type. Network architectures like the number of layers, nodes, and activation functions can be specified on a per-element and per-committee member basis. User-defined atomic reference energies can be subtracted from the training targets. During predictions, these atomic energies are added back to the network output to obtain total energies that are numerically consistent with the underlying reference method.

\subsubsection{Third-Generation HDNNPs}
\label{sec:features_3g}

Local properties needed for computing interactions beyond the cutoff radius (s. Eq.\;\ref{eq:3g}) are predicted by dedicated \textsc{HDNN} instances that may share feature representations with the short-range energy model. The resulting electrostatic and dispersion energies are subsequently evaluated using the interaction models described in Sections~\ref{sec:electrostatics} and~\ref{sec:vdw}; their respective solver implementations are discussed in Section~\ref{sec:impl_electrostatics}. In the code base, a calculator for dispersion interactions is added as a separate type (Fig.~\ref{fig:flowchart} \--- Level 1) and is therefore available for all HDNNP generations and compatible with committee evaluation.

\subsubsection{Fourth-Generation HDNNPs}
\label{sec:features_4g}

In \RuNNer, two property models predict atomic electronegativities $\boldsymbol{\upchi}$ and hardnesses $\mathbf{J}$, respectively. Through the choice of the explicit property model (\textsc{HDNN} or \textsc{ElementalPropertyModel}) it is possible to express these properties as environment- or solely element-dependent quantities. The equilibrium charges are then obtained by minimizing the electrostatic energy under the constraint of total charge conservation, solving the linear system of Eq.~\ref{eq:4G_qeq_prime}. The concrete solver implementations and their computational scaling are discussed in Section~\ref{sec:impl_electrostatics}.

\subsection{Feature Map Implementation}
\label{sec:impl_features}

The feature map framework in \RuNNer is designed for extensibility: new feature maps can be added by implementing a well-defined interface and are then registered in the feature calculator, without the need to modify any other part of the library. \RuNNer ships reference implementations of atom-centered symmetry functions and the overlap matrix descriptor.

\subsubsection{Atom-Centered Symmetry Functions}

To accelerate feature evaluation, ACSFs sharing common parameters, such as the cutoff function, neighbor element specifications, and ACSF type, are grouped into batches to compute multiple features simultaneously, with the per-feature parameters ($\eta$, $R_\text{shift}$, $\zeta$) stored as arrays~\cite{P5603}. Common intermediate quantities such as pairwise distances, angles, and cutoff function values are computed once per group and reused across all members, eliminating redundant calculations. The batch structure also exposes single instruction, multiple data (SIMD) vectorization opportunities across the per-feature parameter arrays. Quantities depending only on the feature parameters but are independent of the atomic positions are precomputed at initialization and cached, avoiding repeated evaluations on the same system.

\subsubsection{Overlap Matrix}
\label{sec:impl_om}

The overlap matrix features (Section~\ref{sec:theory_om}) are implemented as a feature map type following the same interface as the ACSF types. The smooth spectral projection of Eq.~\ref{eq:om_gridmap_runner} is the default variant; the original sorted-eigenvalue form is also available.

\subsection{Electrostatics Solvers}
\label{sec:impl_electrostatics}

For periodic systems, which we consider as the typical scenario, Eq.~\ref{eq:elec_energy} is evaluated via direct Ewald summation~\cite{ewald_berechnung_1921}, scaling as $\mathcal{O}(N^{3/2})$, or via two mesh-based schemes\---a plane-wave solver and PPPM~\cite{essmann_smooth_1995}\---both of which achieve $\mathcal{O}(N\log N)$ scaling by evaluating the Coulomb kernel in reciprocal space via FFT. For non-periodic systems a direct pairwise Coulomb sum is used.

For 4G models, the charges are obtained by solving the $(N+1)\times(N+1)$ system of linear equations given in Eq.~\ref{eq:4G_qeq_prime} at every evaluation step. A direct LU factorization scales as $\mathcal{O}(N^3)$ and is feasible only for small to medium-sized systems. Iterative conjugate gradient (CG) solvers reduce the per-step cost, but a naive matrix-vector product $\mathbf{A}\mathbf{v}$ still scales $\mathcal{O}(N^2)$ because the Coulomb kernel couples every pair of atoms.

For these reasons, \RuNNer employs the quasi-linear-scaling particle mesh charge equilibration scheme of Gubler \textit{et al.}~\cite{gubler_accelerating_2024}. The central insight is that the matrix-vector product $(\mathbf{A}\cdot\mathbf{q})_i$ is equivalent to the gradient of the electrostatic energy with respect to the charge on atom $i$,
\begin{equation}
(\mathbf{A}\cdot\mathbf{q})_i = \frac{\partial E_\text{elec}}{\partial q_i} = \int \rho_i(|\mathbf{r}-\mathbf{r}_i|)\,V(\mathbf{r})\,\mathrm{d}\mathbf{r}\,,
\label{eq:matvec}
\end{equation}
where $\rho(\mathbf{r}) = \sum_j q_j \rho_j(\|\mathbf{r}-\mathbf{r}_j\|)$ is the current total charge density and $V(\mathbf{r})$ its electrostatic potential. For periodic boundary conditions, $V(\mathbf{r})$ is obtained by solving Poisson's equation in Fourier space: the charge density is projected onto a uniform real-space grid, transformed by FFT, multiplied by the reciprocal-space kernel $4\pi/G^2$ (where $G$ denotes the magnitude of the reciprocal lattice vector $\mathbf{G}$), and back-transformed. Evaluating the integral of the electrostatic potential $V(\mathbf{r})$ over the localized atomic density $\rho_i$ then yields $(\mathbf{A}\cdot\mathbf{q})_i$ for every atom at $\mathcal{O}(N\log N)$ total cost, dominated by the FFT; the integral is quasi-local because the Gaussian densities $\rho_i$ decay exponentially, so only a small real-space volume around atom $i$ needs to be sampled. Crucially, this procedure evaluates the matrix-vector product without ever forming the $N\times N$ Coulomb matrix $\mathbf{A}$. The same Fourier-space machinery also yields analytical forces and the stress tensor without increasing the computational scaling complexity, by evaluating further convolutions with the same transformed potential.

In practice, the number of CG iterations often increases with system size. Gubler \textit{et al.} found empirically that for real systems the overall method scales as $\mathcal{O}(N\log^2 N)$. This algorithmic advance is the origin of the orders-of-magnitude higher efficiency compared to other 4G implementations as demonstrated in Fig.~\ref{fig:lammps_performance}.

\subsection{Computational Performance Considerations}

\subsubsection{Order of Feature Evaluation in Training}

Using the force vector of atom $i$ for training does not only require its feature values $\mathbf{G}_i$ but also the partial derivatives with respect to its nuclear coordinates of all feature vectors of the neighboring atoms containing the central atom $i$ in their respective cutoff radii (see Eq.\;\ref{eq:2G_forces}). \RuNNer provides two complementary evaluation strategies, which are selected automatically based on the fraction of force components/vectors per structure batch that are selected for parameter updates. Importantly, these strategies only optimize performance and do not influence the final training results (up to numerical differences caused by altered summation order).

In the \emph{atom-centric} route, all derivatives of a central atom $i$'s feature vector with respect to its neighbors are computed in a single pass and the resulting Jacobian is cached for the whole structure batch. This Jacobian is then reused for every force component in the batch that involves atom $i$ as a neighbor. This route is mandatory for 4G-HDNNPs because the global charge feature creates a dense dependency between every atom's energy and every other atom's position; it is also preferred for small periodic structures where the number of neighbors per atom is larger than the total number of atoms in the unit cell, and when the fraction of forces to be used in training is above 50\%. In the \emph{neighbor-centric} route, only the derivatives of neighbors' feature vectors with respect to one specific target atom are evaluated and no batch-wide Jacobian is stored, which reduces peak memory substantially if only a small fraction of atoms is selected for force updates per batch.

For small periodic systems with many neighbors per atom an additional optimization sums all periodic-image contributions from a central atom to a target force during packing, so that the number of backward passes per force component scales with the number of atoms in the structure rather than the number of neighbors.

\subsubsection{4G-HDNNP Charge Training}
\label{performance:4G_charge_training}

For 4G charge training, Eq.\;\ref{eq:4G_qeq_prime} must be solved repeatedly to obtain the atomic charges $\mathbf{q}$ based on the updated electronegativities $\boldsymbol{\upchi}$ and hardnesses $\mathbf{J}$. Since
\begin{equation}
\mathbf{q'} = -\mathbf{A'^{-1}}\boldsymbol{\upchi}'\,,
\end{equation}
the Coulomb matrix $\mathbf{A}$ and its inverse $\mathbf{A^{-1}}$ can be cached, which reduces this step to a single efficient matrix-vector product. If the hardness is kept fixed throughout training, the Coulomb matrices for all structures can be precomputed and inverted once before the training loop begins. If the hardness is learned along with the electronegativities, the Coulomb matrix is stored without the hardness contributions to its diagonal elements; after each forward pass, it is then updated with the new hardness values and re-inverted. Even in this case, a substantial performance gain is achieved by avoiding the repeated construction of the Coulomb matrix from scratch.

\subsubsection{HDNNP Committees}
\label{sec:committees}

Committee models enable active learning protocols and often provide more stable predictions through averaging~\cite{P5814}. They consist of $N_\mathrm{comm}$ independently initialized and trained networks that are evaluated jointly for each input structure. Training multiple committee members is supported natively in \RuNNer: each member is trained as an independent network with its own random weight parameter initialization and architecture and the parallelization and memory management strategies described above apply equally to committee training and to single-network training. When features and their derivatives are evaluated on the fly, this reduces the per-committee training cost substantially. Otherwise, the per-member training cost is essentially identical to that of a single model. At evaluation time, \RuNNer computes neighbor lists and feature maps once per call; each of the $N_{\mathrm{comm}}$ networks then performs its own forward pass over the shared feature vectors. The marginal cost of adding a committee member is therefore approximately equal to the cost of a single neural-network forward pass for each member and atom, which is typically small compared to feature evaluation and neighbor-list construction for realistic system sizes. All per-member outputs \---- energies, forces, stresses, charges, and Hirshfeld volumes \---- are stored separately and can be accessed by the user. Mean values are used for MD propagation, while the standard deviation provides a direct measure of uncertainty that can be used to trigger retraining in active learning loops.

\subsection{Available Optimizers}

For optimizing the neural network weights, \RuNNer supports gradient descent (GD), stochastic gradient descent (SGD), Adam~\cite{P5404}, and two variants of the extended Kalman filter (standard and fading-memory)~\cite{P1308}. The gradient-based optimizers (GD, SGD, and Adam) are parallelized by averaging gradients across MPI ranks after each local backward pass. The Kalman filter variants require a different procedure because the update equations couple all parameters through a covariance matrix: for these, the multistream formulation of Singraber \textit{et al.}~\cite{singraber_parallel_2019} is used, in which each MPI rank provides one data stream and the full Kalman update is performed collectively. This makes the Kalman filter scale under MPI parallelism nearly as efficiently as gradient descent. Mathematical details are provided in the Supporting Information, Sec.~\ref{section:si:training:opt:kalman}.

\subsection{Interfaces to External Software}
\label{sec:interfaces}

\RuNNer provides C binding interfaces to key computational routines, enabling straightforward integration into other programming language environments and simulation codes. In the following, this is demonstrated for native interfaces to LAMMPS~\cite{P4473} and Python, the latter being further integrated into the Atomic Simulation Environment (ASE)~\cite{P5915}.

\subsubsection{LAMMPS Interface} \label{sec:lammps}

\RuNNer provides a native pair style for LAMMPS~\cite{P4473} supporting all HDNNP generations\---including 4G charge equilibration, dispersion corrections, and committee uncertainty quantification\---within a standard LAMMPS simulation workflow. The interface is implemented as a thin C++ layer on the LAMMPS side that communicates with the \RuNNer shared library through ISO C bindings. This design decouples the two codebases: the \RuNNer library evolves independently, and the narrow C interface only requires updates when new library functionalities need to be exposed to the driver application.

The neighbor list produced by LAMMPS' domain decomposition is converted into \RuNNer's internal format before feature evaluation, supporting multiple simultaneous cutoff radii. For 3G and 4G models, long-range electrostatics and QEq require a globally consistent view of the atomic charges: the interface uses LAMMPS's forward and reverse communication layers to assemble the full local-plus-ghost atom list before QEq and to distribute the resulting atomic charges back to the owning ranks afterwards. Stress tensors are computed through the virial theorem for 2G-HDNNPs and analytically for all other HDNNP generations and passed back to LAMMPS in the standard format required for $NPT$ simulations.

Committee models are evaluated in a single call to the \RuNNer library, with the averaged energies and forces used for MD propagation. Individual member predictions are optionally stored as per-atom arrays, providing direct access to uncertainty estimates during production simulations without any additional computational overhead. The interface also provides feature extrapolation detection: at each MD step a check is performed if any atomic environment falls outside the training data distribution, which can be used as a starting point for active learning or to terminate the simulation before unphysical behavior might occur.

Implemented as a regular pair style,\RuNNer supports many LAMMPS packages, e.g. its connection to the enhanced sampling engine PLUMED~\cite{bonomi_plumed_2009}, by design.

\subsubsection{ASE: \textsc{RuNNerlib}} \label{sec:ase_interface}

\RuNNer provides a Python calculator for the Atomic Simulation Environment (ASE)~\cite{P5915} that enables access to the full suite of \RuNNer capabilities from within ASE's ecosystem. The key design choice is to communicate directly with the \RuNNer shared library through Python's \texttt{ctypes} interface rather than through on-disk files: atomic positions, cell vectors, and species are passed in memory and energies, forces, and stresses are returned in the same call. This eliminates the I/O overhead that makes file-based calculators less performant, especially for large-scale simulations. The plugin is distributed as a standalone package~\cite{runner-interface_2026} that requires only a working \RuNNer installation; no modifications to ASE are needed.
Committee models are supported in ASE in an identical way to the LAMMPS interface, with averaged properties used by default and individual member outputs available for uncertainty estimation.

\subsubsection{ASE: \textsc{RuNNerbin} and Workflow Management} \label{sec:ase_workflows}

The \textsc{RuNNerlib} calculator described above is optimized for low-latency in-memory evaluation and is the appropriate interface for molecular dynamics, geometry optimization, and on-the-fly uncertainty monitoring. Managing a complete potential development workflow including dataset curation, feature design, iterative training, convergence analysis, and active learning, requires a different integration point. This is provided by the broader \runnerase ecosystem, a collection of Python packages that expose the full \RuNNer workflow through ASE-compatible interfaces.

At the workflow level, \textsc{RuNNerbin} is a subprocess-based calculator that invokes the \RuNNer executable directly and manages the associated file-based protocols for training and standalone prediction. Rather than communicating through the shared library in memory, \textsc{RuNNerbin} prepares the required input files, dispatches the binary, and parses all output back into structured Python objects. This design is intentional: file-based decoupling enables the restart of training jobs, monitoring, and compatibility with batch schedulers, where the Python orchestration process and the compute job run independently. A lightweight SLURM submission utility included in the ecosystem allows submitting complex training workflows as sequences of dependent batch jobs, with states managed between individual runs entirely in Python.

Every quantity produced by a \RuNNer training run is represented by a dedicated storage class that reads and writes \RuNNer's native file formats and exposes the data as NumPy arrays. This includes network weights, input and output scaling parameters, precomputed feature values, dataset partitions, and per-epoch metrics including energies, forces, stresses, and arbitrary atomic properties like charges or Hirshfeld volumes. This makes the complete training state directly accessible to Python-based postprocessing without manual file parsing. Feature maps can be generated programmatically from physical heuristics and pruned by variance or importance thresholds derived from the training data, with separate utilities for dataset inspection, composition analysis, and structure visualization with OVITO~\cite{ovito}.

The \runnerase ecosystem also provides a self-contained active learning framework that couples committee disagreement directly to the training loop. During a simulation, each evaluation is processed by all committee members; when the spread of their predictions exceeds a configurable threshold, the flagged configuration is collected, submitted for DFT evaluation, and incorporated into the training set before the next training round. Because committee support runs through all \RuNNer workflows without additional overhead, this automated retraining cycle requires no modifications to the simulation protocol.

The reader and writer for \RuNNer's native structure format are maintained as a part of the official ASE distribution and are available to ASE users without additional installation. All other \runnerase packages are distributed as independent entries\cite{runnerase-core_2026,runnerase-prediction_2026,runnerase-training_2026,runnerase-workflows_2026,runner-interface_2026,runnerase-feature-calculation_2026} and a single wrapper package~\cite{runnerase_2026} on the Python Package Index (PyPI), allowing users to conveniently install all packages at once or only the components relevant to their use case.

\section{Computational Performance}\label{sec:performance}

We benchmark \RuNNer's computational performance on CPU-based hardware for three typical applications: molecular dynamics simulations in LAMMPS~\cite{P4473}, property evaluation via the ASE interface~\cite{P5915}, and model training using \RuNNer directly. LAMMPS benchmarks are compared against the open-source HDNNP codes n2p2~\cite{P5603} and PANNA~2.0~\cite{P6532}. Timings are reported for large supercells of a water reference system (96 atoms per unit cell) at the respective compiler-optimized production settings. Full details of the test systems are provided in the SI, Sec.~\ref{sec:si:methods:perf_bechmarks}.

\subsection{LAMMPS MD Performance}

\begin{figure*}[!htbp]
    \centering
    \includegraphics[width=\linewidth]{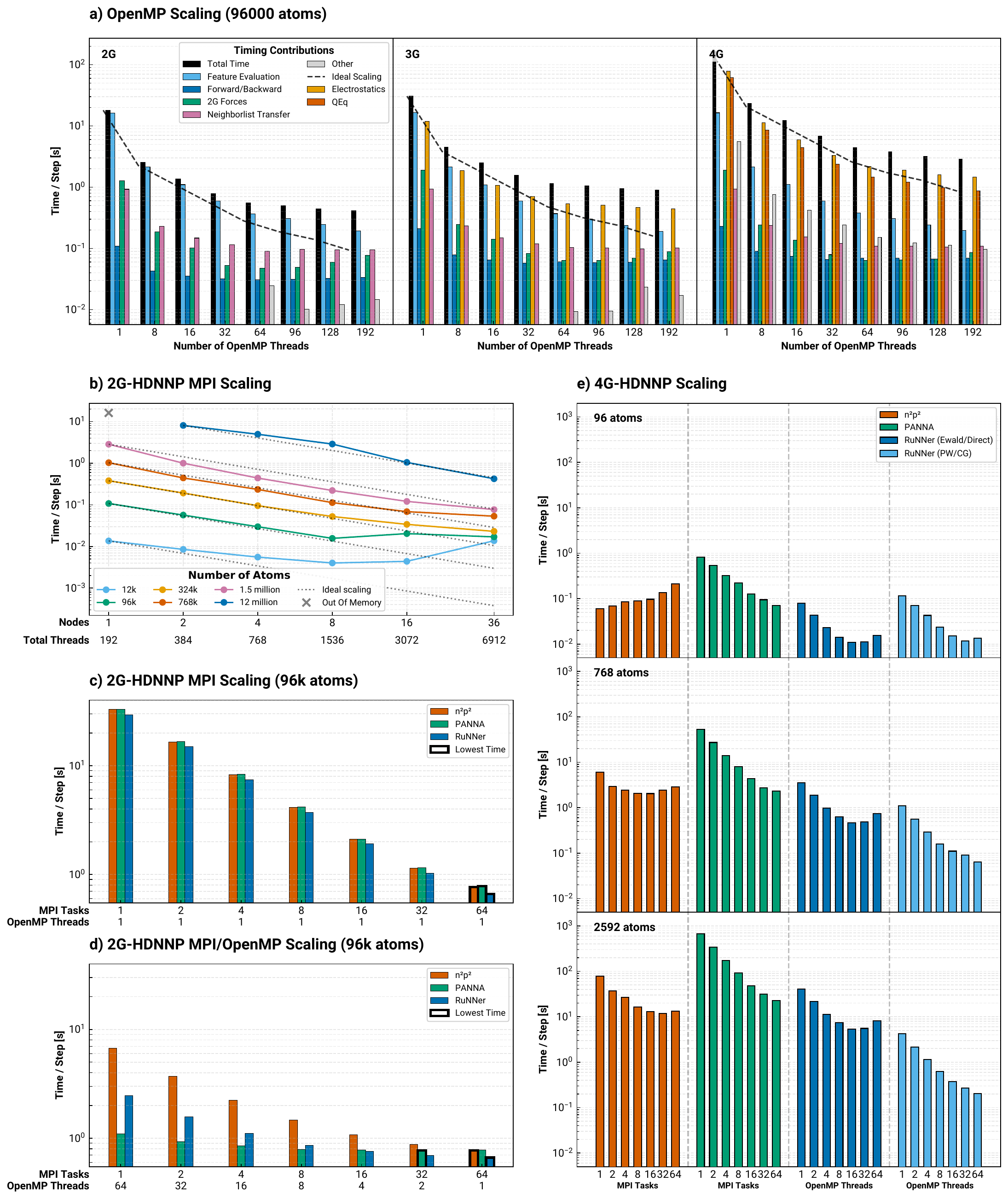}
    \caption{ Prediction performance of \texttt{pair\_style runner} in LAMMPS for a water reference system. \textbf{(a)}~OpenMP thread scaling for 2G, 3G, and 4G-HDNNPs at 96,000 atoms, with per-step wall time decomposed by timing contributions. \textbf{(b)}~2G MPI strong scaling for 2G-HDNNPs from 12,000 to about 12 million atoms for up to 36 compute nodes (192--6912 total threads); the marker labeled OOM indicates an out-of-memory failure. \textbf{(c)}~Pure-MPI scaling at 96,000 atoms comparing \RuNNer, n2p2~\cite{P5603}, and PANNA~\cite{P6532}. \textbf{(d)}~Hybrid MPI/OpenMP scaling at 96,000 atoms for the same three codes. \textbf{(e)}~4G-HDNNP scaling for 96, 768, and 2592 atoms as a function of MPI tasks (left two columns) and OpenMP threads (right two columns), comparing \RuNNer (Ewald/direct and plane-wave/CG), n2p2, and PANNA. Detailed test setup and hardware specifications are given in the SI, Secs.~\ref{sec:si:methods:perf_bechmarks:lammps} and \ref{sec:si:methods:comparison}.}
    \label{fig:lammps_performance}
\end{figure*}

Figure~\ref{fig:lammps_performance} summarizes the performance of the \texttt{pair\_style runner} for energy and force predictions across different system sizes, parallelization strategies, and HDNNP generations. Detailed hardware specifications are given in the SI, Sec.~\ref{sec:si:methods:perf_bechmarks}.

All generations exhibit favorable scaling with the number of OpenMP threads (Fig.~\ref{fig:lammps_performance}a) for a 96,000-atom system. For 2G-HDNNPs, the runtime is dominated by feature evaluation, with minor contributions from force calculation ($\frac{\mathrm{d}E}{\mathrm{d}G}\frac{\mathrm{d}G}{\mathrm{d}r}$) and neighbor list assembly. Near-ideal speedup is maintained up to 64 threads, beyond which synchronization overhead limits further gains. Adding electrostatics in the 3G-HDNNP case introduces a major contribution from the long-range solver in the same order of magnitude as the feature evaluation. The 4G-HDNNP model carries the additional cost of the global charge equilibration; consequently, the runtime is dominated by QEq and electrostatic force evaluation via the force trick (see Sec.\;\ref{sec:electrostatics}). The FFT operations in the plane-wave solver exhibit near-ideal scaling up to 96 threads, at which point memory access across the two CPUs of a single compute node becomes a bottleneck. Nevertheless, the code still benefits from additional resources up 192 threads, i.e., the full compute node, with a maximum 4G-HDNNP throughput of approximately \SI{0.03}{\nano\second\per\day} based on a time step of 1~fs.

For 2G-HDNNPs, MPI strong scaling stays ideal for up to 36 compute nodes (6,912 cores) across all investigated system sizes from 12,000 to approximately 12 million atoms, with no evidence of a communication bottleneck until the per-node workload becomes too small in comparison to MPI overhead (Fig.~\ref{fig:lammps_performance}b). For instance, simulation length per day for 1.5 million atoms increases from 0.03 ns/day (\SI{1}{\femto\second} timestep) on a single node, to 1.12 ns/day when using all 36 nodes. The single out-of-memory point for 12 million atoms when using only one node (total RAM: 768 Gigabyte) confirms that memory, not performance, is the practical constraint at this scale. In all cases, optimal performance is reached by exploiting \RuNNer's hybrid parallelization, assigning eight OpenMP threads to each MPI task. As evident in panels~(c) and~(d) the 2G-HDNNP performance of \RuNNer is on par with the HDNNP implementations n2p2~\cite{P5603} and PANNA~\cite{P6532} on a 96,000-atom system under pure-MPI and hybrid MPI/OpenMP parallelization, respectively. \RuNNer reaches lower absolute wall times than both codes across the tested configurations, with the gap widening at low MPI task counts where \RuNNer's more efficient per-core feature evaluation provides the largest advantage. It is noted that these tests are not exhaustive, and all three codes perform competitively. Nevertheless, this result is particularly striking when considering the design philosophy of the different frameworks. \RuNNer focuses strongly on reusable workflows, employing a unified codebase where the LAMMPS interface utilizes the exact same high-performance prediction routines as the ASE interface and the standalone \RuNNer training binary.

In contrast, other implementations often diverge at the architectural level; for instance, PANNA maintains a fully independent LAMMPS interface, a design choice that facilitates its excellent OpenMP parallelization characteristics. Furthermore, \RuNNer's development was fundamentally centered around the complex challenges of 4G-HDNNPs, where global charge equilibration rather than local potential evaluation represents the primary computational bottleneck. The fact that \RuNNer maintains such high efficiency for 2G-HDNNPs despite this focus on 4G-specific optimizations underscores the robustness of its core kernels.

Demonstrably, \RuNNer outperforms both n2p2 and PANNA in 4G-HDNNP simulations of systems with 96, 768, and 2592 atoms (Fig.~\ref{fig:lammps_performance}e). Simulations of larger systems are possible with \RuNNer (Fig.~\ref{fig:lammps_performance}a), whereas the computational scaling of the other codes limits their use to moderately sized systems. For each code, we choose the optimally supported parallelization scheme (MPI for n2p2 and PANNA, OpenMP for \RuNNer). The n2p2 4G timings use the iterative QEq implementation of Kocer \textit{et al.}~\cite{kocer_iterative_qeq_2025}, with an improved solver contributed in preparation for this benchmark;\cite{knoll_n2p2_pr_2024} even with these improvements, n2p2 is slower than \RuNNer by up to two orders of magnitude, as is PANNA. We note that PANNA uses a slightly different ACSF implementation and predicts the required atomic properties as part of a single model with multiple output nodes. The fundamental reason is algorithmic: n2p2's iterative QEq requires one full $\mathcal{O}(N^2)$ matrix--vector product with the dense Coulomb matrix per CG iteration, whereas \RuNNer's plane-wave decomposition reduces each such product to $\mathcal{O}(N\log N)$. As system size grows from 96 to 2592 atoms the gap widens accordingly. In \RuNNer, the plane-wave/CG variant consistently outperforms the direct Ewald solver confirming that the reciprocal-space decomposition is the more practical choice for all but the smallest simulation cells.

\subsection{RuNNerlib ASE Performance}

\begin{figure}[!htbp]
    \centering
    \includegraphics[width=.8\columnwidth]{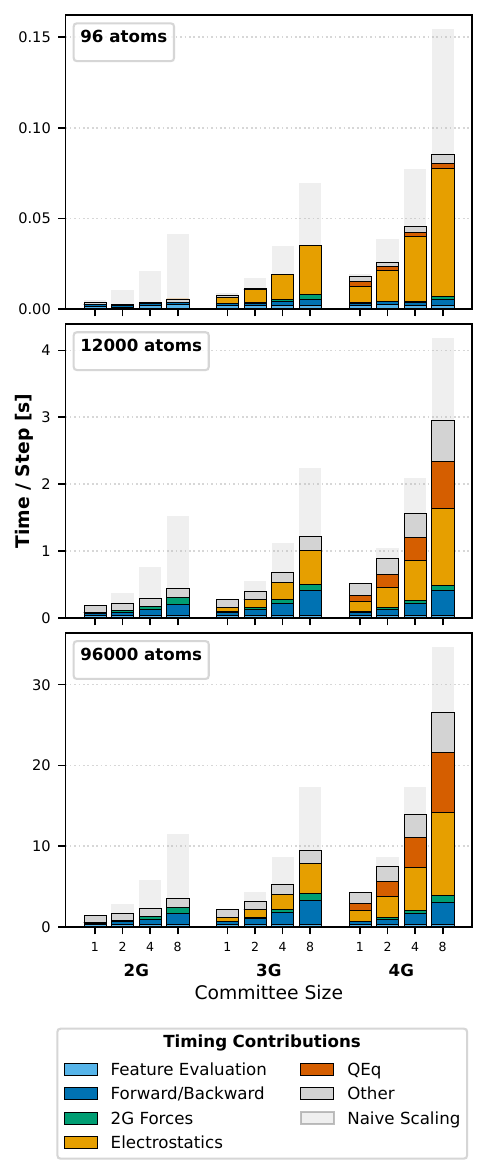}
    \caption{ Wall time per evaluation step of the \textsc{RuNNerlib} ASE calculator as a function of committee size for 2G-, 3G-, and 4G-HDNNPs at three system sizes (96, 12,000, and 96,000 atoms). Bars are decomposed by timing contribution; gray bars indicate the naive overhead expected from independent sequential committee evaluations. Detailed hardware specifications are given in the SI, Sec.~\ref{sec:si:methods:perf_bechmarks:ase}.}
    \label{fig:ase_performance}
\end{figure}

The ASE calculator interface (\textsc{RuNNerlib}) provides a Python-accessible route to \RuNNer potentials for, e.g., structure optimizations, vibrational analysis, and molecular dynamics~\cite{P5915}. Figure~\ref{fig:ase_performance} shows the wall time as a function of committee size for three system sizes (96, 12,000, and 96,000 atoms) and all three HDNNP generations.

The actual wall time grows far more slowly than the naive expectation of independent committee member evaluations (gray reference bars). Because neighbor lists and feature vectors are computed once and shared across all committee members, the cost of adding a member is limited to the neural network forward pass itself, which is a small fraction of the total step time for realistically sized systems. Committees of four to eight members are therefore computationally affordable even for 96,000 atoms, which matters in practice given their role in uncertainty quantification and active learning. As expected from the LAMMPS benchmarks, the contribution of the electrostatic and QEq contributions to the total cost grows with increasing system size and generation. For a 4G-HDNNP model with a committee size of eight at 96,000 atoms, solving the QEq equations accounts for the majority of the per-step time, while for 2G-HDNNP models the forward pass remains the dominant contribution at all sizes tested.

\subsection{Training Performance}

Models trained using \RuNNer reach state-of-the-art accuracy on energy ($\leq1~\mathrm{meV\;atom}^{-1}$) and force labels ($\leq100~\mathrm{meV\;\AA}^{-1}$). Apart from that, efficient training is critical for the iterative workflows underpinning HDNNP development, since active learning, hyperparameter optimization, and committee construction require many training runs. 

\begin{figure}[!htbp]
    \centering
    \includegraphics[width=\columnwidth]{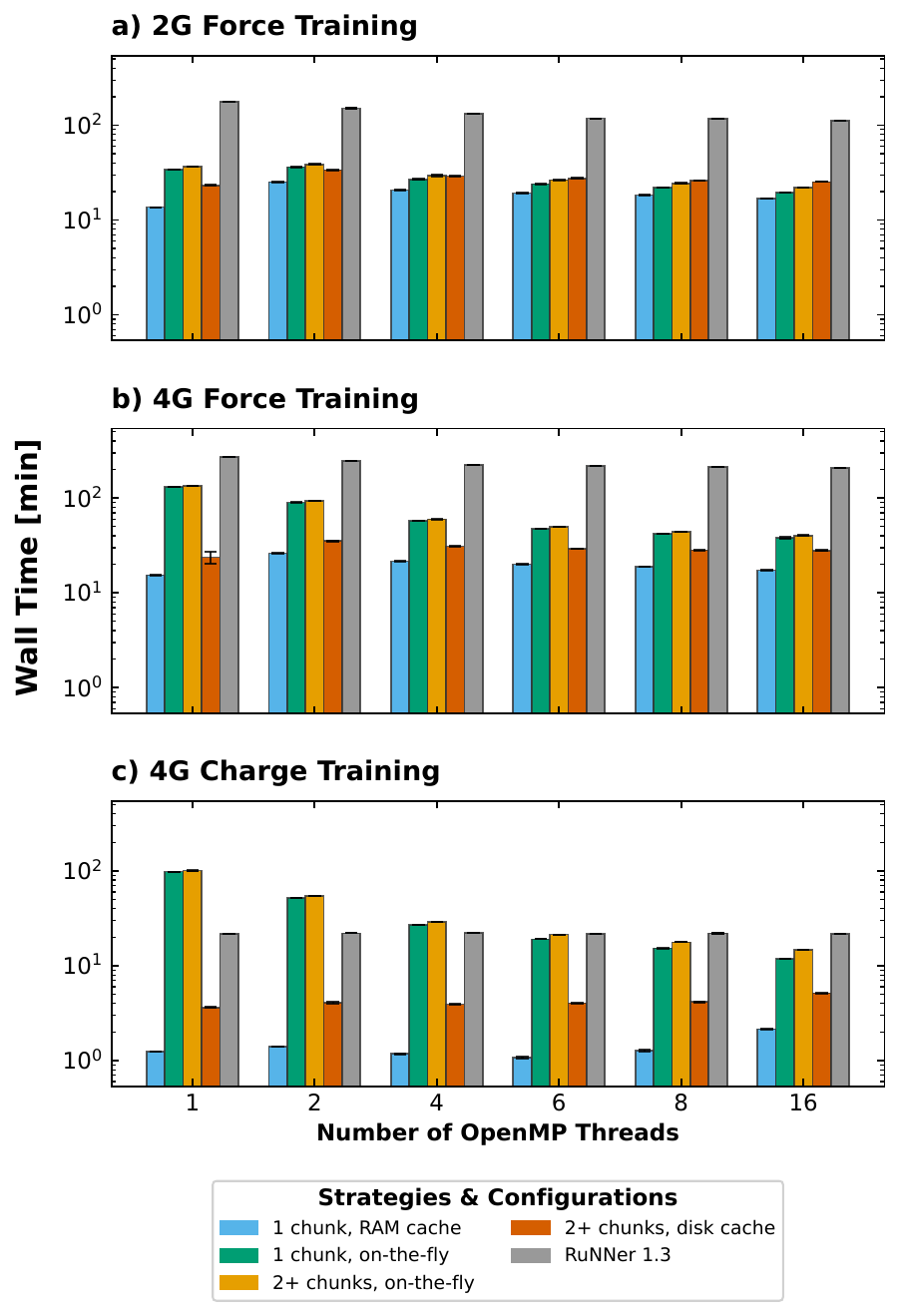}
    \caption{Total training wall times for 100 epochs using \RuNNer versus \textsc{RuNNer}~1.3 as a function of OpenMP thread count for four memory management strategies: full RAM cache (one chunk, precomputed), on-the-fly evaluation with one chunk, on-the-fly with multiple chunks, and disk cache. \textbf{(a)}~2G force training. \textbf{(b)}~4G force training. \textbf{(c)}~4G charge network training. Details about the training data and test setup are provided in the SI, Sec.~\ref{sec:si:methods:training}.}
    \label{fig:training_performance}
\end{figure}

Figure~\ref{fig:training_performance} compares total wall times for 100 training epochs with \RuNNer with its predecessor RuNNer~1.3 employing four memory management strategies as a function of OpenMP thread count. The test setup was chosen such that both codes achieve similar accuracy with a nearly identical number of optimizer steps. Further details are provided in the SI, Sec.~\ref{sec:si:methods:training}.

The four strategies span the spectrum from maximum throughput on a HPC cluster to limited resources, e.g., on a local workstation. When the entire dataset fits in RAM, all features and their derivatives are precomputed once and cached before the loop over epochs begins, keeping feature evaluation entirely outside the inner loop. When memory is more limited, \RuNNer may load the dataset in multiple chunks, in which case the features can be recomputed on-the-fly for each structure batch and discarded immediately, or the precomputed features can be read from a disk cache. For comparison purposes, we additionally test the intermediate case of loading the full dataset in a single chunk, but computing features on-the-fly nonetheless. The choice of strategy has significant impact on all training procedures, but especially so on 4G charge and force training, because 4G models require storing global charge features and the Coulomb matrix in addition to the standard per-atom feature derivatives.

For 2G-HDNNP force training (Fig.~\ref{fig:training_performance}a), \RuNNer is an order of magnitude faster than \textsc{RuNNer~1.3} at identical thread counts. Both the RAM-cached and disk-cached variants of \RuNNer show similar absolute wall times and negligible sensitivity to thread count. This is because the training loop is dominated by the optimizer step of the Kalman filter, which profits little from OpenMP parallelization. This provides an impressive lower bound for the overall training time with \RuNNer. However, a fair comparison to the RuNNer~1.3 baseline must take into account that the older code recomputes all feature derivatives on-the-fly. In this low-memory scenario, \RuNNer still significantly outperforms the preceding code version. This is mainly due to the strongly optimized feature evaluation in \RuNNer. Since 3G-HDNNP force training is algorithmically analogous to 2G training, we do not show explicit timings.

For 4G-HDNNP force training (Fig.~\ref{fig:training_performance}b), the advantage of full RAM caching is most pronounced: with all (global) features and their derivatives precomputed, \RuNNer completes each epoch 25 times faster than \textsc{RuNNer~1.3}. Notably, the 4G force training time in \RuNNer is only negligibly slower than the 2G-HDNNP force training time.
On-the-fly feature computation narrows the advantage because the global charge feature derivatives are expensive to evaluate. However, this setup profits most from OpenMP parallelization, bringing it down to the same order of magnitude as the cached variant at 16 threads. For 4G-HDNNP charge training (Fig.~\ref{fig:training_performance}c), \textsc{RuNNer~1.3} precomputes the Coulomb matrix of all structures before beginning the training loop. The comparison with this strategy in \RuNNer exhibits a five- to tenfold speedup. Notably, \RuNNer is still faster than \textsc{RuNNer~1.3} re-evaluating the Coulomb matrix on-the-fly when executed in OpenMP-parallel.

\section{4G-HDNNP Benchmark: Non-Local Charge Transfer in A\lowercase{u}/M\lowercase{g}O}


In order to show in which situations 4G-HDNNPs are required to obtain a physically correct PES that cannot be provided by local or semilocal MLPs, we use an adapted version of the benchmark originally designed by Ko~\textit{et al.} and investigate the adsorption of a gold adatom on a clean and aluminum-doped MgO(001) slab formed by a $3\times3$ supercell. To probe the range dependence of the effect, we consider a series of three slab thicknesses spanning 3, 4, and 5 bulk unit cells ($3\times3\times3$, $3\times3\times4$, and $3\times3\times5$ supercells).

On the pristine MgO(001) surface, a gold adatom interacts primarily through electrostatic polarization and dispersion, with a modest energetic preference for adsorption above an oxygen atop site. The character of this interaction changes fundamentally when a deep subsurface Mg$^{2+}$ ion is replaced by Al$^{3+}$. In DFT, the excess positive charge of the Al$^{3+}$ dopant in combination with the constraint of an overall neutral system results in a negatively charged Au atom.
In response, the two adsorption sites become almost iso-energetic, with a slight preference for the magnesium atop site in case of the $3\times3\times3$ slab. This charge redistribution extends over many lattice spacings, i.e., the charge state of the adsorbed Au atom is sensitive to a dopant that is located well outside the receptive field of the HDNNP.
Local MLPs like 2G-HDNNPs, but also MLPs including long-range electrostatics based on environment-dependent charges only, cannot represent this effect and result in a qualitatively incorrect PES.
Since this charge transfer is present irrespective of the distance between the Al and the Au atoms, also message passing-based MLPs are generally unable to capture this charge transfer as it is always possible to construct a slab that is thicker than the range covered by the message passing steps.

To demonstrate the need for a global description, we trained both a 2G- and a 4G-HDNNP committee on energies only, using the same DFT reference dataset, consisting of adsorption energy curves as a function of Au--surface distance for adsorption above the O and Mg atop sites in both a pristine and an Al-doped MgO slab across all three thicknesses ($3\times3\times3$, $3\times3\times4$, and $3\times3\times5$ supercells).

\begin{figure*}[!htbp]
    \centering
    \includegraphics[width=0.85\linewidth]{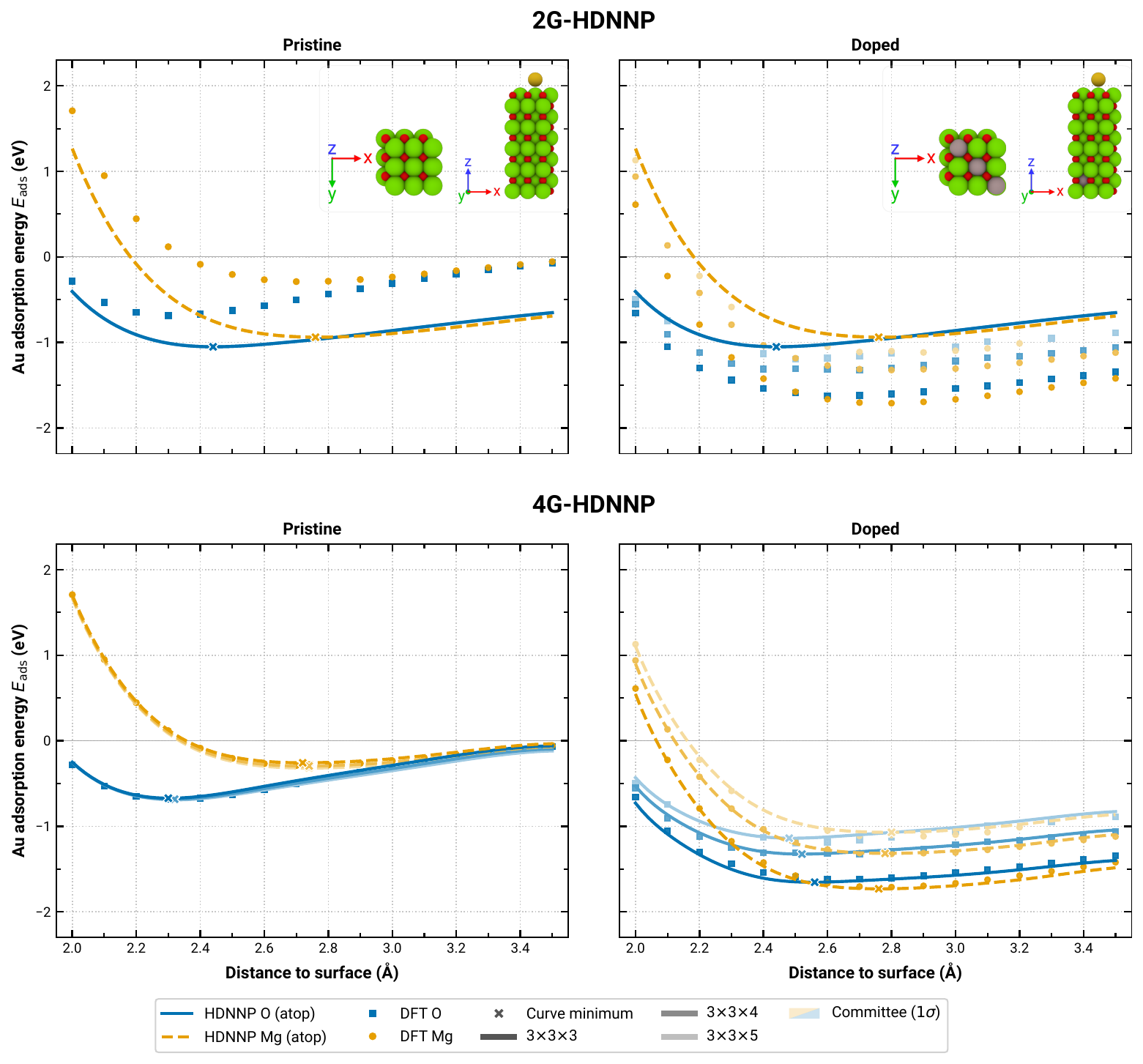}
    \caption{Adsorption energy $E_\mathrm{ads}$ of a gold adatom on a MgO(001) surface as a function of the vertical Au--surface distance for adsorption above the O (blue/solid) and Mg (orange/dashed) atop sites computed with a local 2G-HDNNP committee (upper row) and a global 4G-HDNNP committee (lower row). The energy is referenced to the isolated Au atom and the bare slab, so that $E_\mathrm{ads}=0$ corresponds to the Au atom far from the surface. The left column shows the pristine MgO surface, while the right column refers to the doped surface with three Mg$^{2+}$ replaced by Al$^{3+}$ at the bottom of the slab far from the adsorption site. Within each panel, three MgO(001) slab thicknesses ($3\times3\times3$, $3\times3\times4$, and $3\times3\times5$ supercells) are overlaid and encoded by color shade from dark (thick, $3\times3\times3$) to light (thin, $3\times3\times5$). The HDNNP predictions (colored lines) are compared to the DFT reference (small markers); for the doped $3\times3\times5$ slab a few DFT single points did not reach SCF convergence and are omitted from those reference curves. Crosses denote the HDNNP energy minimum of each curve. Insets in the upper (2G) row show top ($z$) and side views of the pristine (left) and doped (right) slab.}
    \label{fig:4g_benchmark}
\end{figure*}

As expected, the local 2G-HDNNP model produces exactly identical adsorption curves for the pristine and doped slabs at every thickness (Fig.~\ref{fig:4g_benchmark}, upper row), being blind to the deep subsurface dopant, and a geometry optimization results in the adsorption of the Au atom at an oxygen site in both cases. We note that the exact energy landscape learned by the 2G-HDNNP in principle represents an average across the two different systems and the exact shape will vary depending on the training run aiming to minimize the overall loss.

On the other hand, the 4G-HDNNP model, whose QEq scheme enables charges to respond to the global structure of the slab, matches the DFT reference for both the pristine and doped cases with high fidelity at all three slab thicknesses, correctly capturing the doping-induced change in both the binding strength and the site preference. Since the effect is reproduced independently of slab thickness, this directly illustrates why message passing-based MLPs cannot capture it either: one can always construct a slab thicker than the message-passing range.
The committee disagreement remains narrow throughout the well-sampled distance range and widens only close to the repulsive wall where training data are sparse.

This benchmark demonstrates that the limitations of local models in charge-transfer systems are not merely quantitative but categorical, and that the 4G-HDNNP framework implemented in \RuNNer overcomes these limitations without sacrificing transferability to the non-charge-transfer regime.

\section{Conclusions}

We have presented \RuNNer, a high-performance software suite for training and applying second-, third-, and fourth-generation high-dimensional neural network potentials. \RuNNer has been designed to run on a wide range of cost-effective CPU-based architectures, from simple laptops via desktop workstations to large and massively parallel HPC environments utilizing thousands of cores across dozens of nodes. 

While large-scale simulations of local MLPs like 2G-HDNNPs have become routine in recent years, to date implementations of global fourth-generation MLPs have remained computationally significantly more demanding. Therefore, a central focus of the development of \RuNNer has been on making large-scale 4G-HDNNP molecular dynamics simulations computationally tractable by overcoming the prohibitive $\mathcal{O}(N^3)$ scaling of QEq with system size. \RuNNer overcomes this bottleneck by using a quasi-linear plane-wave QEq scheme, which evaluates the Coulomb matrix--vector product via FFT at $\mathcal{O}(N \log^2 N)$ cost without ever forming the dense matrix. In practice, this reduces the per-step wall time of 4G-HDNNP molecular dynamics by several orders of magnitude compared to previous implementations making production simulations of large systems feasible. Consequently, \RuNNer achieves linear or quasi-linear scaling with system size across all HDNNP generations. For 2G-HDNNPs, MPI strong scaling in LAMMPS-based MD simulations holds ideally across all tested system sizes up to 12 million atoms and 36 compute nodes comprising approximately 7000 cores.

Full committee support is included in \RuNNer at no structural overhead: neighbor lists and feature vectors are computed once and shared across all members, ensuring that uncertainty estimates come cheaply for inference and enabling on-the-fly quality assessment during production runs.

Similar algorithmic improvements apply to HDNNP training, where these shared committee structures significantly streamline ensemble optimization. Moreover, advanced caching strategies reduce the per-epoch wall time by a factor of 5--25 relative to \textsc{RuNNer~1.3}, successfully closing the training cost gap for energy and force training between 2G- and 4G-HDNNPs.

All performance gains rest on a four-level modular architecture implemented in modern Fortran with hybrid MPI/OpenMP parallelization and tested via a thorough continuous integration pipeline. This clean design ensures that \RuNNer is not only exceptionally robust but also easily extensible for future potential generations and custom features.

To seamlessly support the broader computational chemistry ecosystem, \RuNNer features direct, high-performance interfaces to common simulation environments such as LAMMPS and ASE; for instance, the \textsc{RuNNerlib} in-memory ASE calculator exposes the underlying high-performance kernels to Python with negligible interface overhead. The complete, open-source \RuNNer software suite, including pre-compiled conda packages, user and developer documentation, and application guides, is openly accessible via our public repository.

In summary, \RuNNer closes the gap between the physical fidelity of fourth-generation models and the computational demands that large-scale atomistic simulations impose, opening the 4G-HDNNP framework to systems and timescales that were previously accessible only to local MLPs.

\section{Data and Software Availability}
The source code of \RuNNer is publicly available under a GPL-v3 license at \url{https://gitlab.com/runner-suite/runner2}. Pre-built binaries are provided as conda packages at \url{https://anaconda.org/conda-forge/runner}. The source code of runnerase is hosted on Gitlab at \url{https://gitlab.com/runner-suite/runnerase-2.0} and all packages provided under that namespace can be installed via pip. Training datasets and analysis scripts to produce the figures in this work can be found at \url{https:/gitlab.com/runner-suite/runner2-paper-data}.

\section{Supplementary Material}

See the supplementary material for a comprehensive mathematical account of RuNNer 2.0, details regarding dataset generation and model training, and descriptions of the performance benchmark setups alongside additional benchmark results.

\begin{acknowledgments}
We thank Djamil Maouene, Rainer Oswald, Jasper Kr\"ahe, Bernadeta Prus, Daniel Trzewik, Jan Elsner, Amir Omranpour, Maite B\"ohm, Yeliz Gürdal, and Tsz Wai Ko for careful validation and helpful discussions. Moreover, we thank the Paderborn Center for Parallel Computing (PC2) for the provided computational resources, and Robert Schade for his generous technical support for the benchmark calculations. We thank Aidan Thompson and Axel Kohlmeyer for their support in making \textsc{pair\_style runner} available in LAMMPS, as well as Ask Hjorth Larsen and Adam Jackson for helpful discussion and deployment work on the ASE interface.
This work was supported by the Deutsche Forschungsgemeinschaft (DFG) through the Cluster of Excellence RESOLV (EXC~2033, project number~390677874) and the Priority Programme SPP~2363 ``Utilization and Development of Machine Learning for Molecular Applications'' (project number~495842446).
A.L.M.K.\ gratefully acknowledges support from the Studienstiftung des deutschen Volkes.
\end{acknowledgments}


\onecolumngrid 
\clearpage

\setcounter{equation}{0}
\setcounter{figure}{0}
\setcounter{table}{0}
\setcounter{page}{1}
\setcounter{section}{0}

\renewcommand{\theequation}{\arabic{equation}}
\renewcommand{\thefigure}{\arabic{figure}}
\renewcommand{\thetable}{\arabic{table}}
\renewcommand{\thesection}{\arabic{section}}

\begin{center}
    {\large \bfseries Supplementary Information: \RuNNer: A Software Suite for High-Dimensional Neural Network Potentials} \\[2.5ex]
    
    \begin{minipage}{0.75\textwidth}
    \begin{center}
    {Alexander L. M. Knoll,$^{1,2}$ Moritz R. Sch\"afer,$^{1,2}$ K. Nikolas Lausch,$^{1,2}$ Moritz Gubler,$^{3,4}$ Henry Wang,$^{1,2}$, Richard Springborn,$^{1,2}$ Redouan El Haouari,$^{1,2}$ Alea Miako Liebetrau,$^{1,2}$ Jonas A. Finkler,$^{3}$,
    Emir Kocer,$^{1,2}$ Marco Eckhoff,$^{5}$ Stefan Goedecker,$^{3}$ Gunnar Schmitz,$^{1,2}$ and J\"{o}rg Behler$^{1,2,*}$}\\[2ex]

    \small
    $^{1}$\textit{Lehrstuhl f\"ur Theoretische Chemie II, Ruhr-Universit\"at Bochum, 44780 Bochum, Germany} \\
    $^{2}$\textit{Research Center Chemical Sciences and Sustainability, Research Alliance Ruhr, 44780 Bochum, Germany} \\
    $^{3}$\textit{University of Basel, Department of Physics, Klingelbergstrasse 82, CH-4056 Basel, Switzerland} \\
    $^{4}$\textit{PSI Center for Scientific Computing, Theory and Data, Paul Scherrer Institute, 5232 Villigen PSI, Switzerland} \\
    $^{5}$\textit{ETH Zurich, Department of Chemistry and Applied Biosciences, Vladimir-Prelog-Weg 2, 8093 Zurich, Switzerland} \\[1ex]
    \end{center}
    \end{minipage}
    
    \vspace{1.5cm} 
\end{center}

\section{Theory}

\subsection{Atomic Features}
\label{section:si:features}

\RuNNer implements two main classes of mappings from an atom's local chemical environment to a fixed-length feature vector: atom-centered symmetry functions (ACSFs)~\cite{P2882} and the overlap matrix (OM)~\cite{sadeghiMetricsMeasuringDistances2013,parsaeifardAssessmentStructuralResolution2021}.
Both classes depend on a smooth cutoff function that enforces strict locality by suppressing contributions from atoms beyond a chosen cutoff radius $r_{\mathrm{cut}}$.

\subsubsection{Cutoff Functions}
\label{section:si:features:cutoffs}

All feature maps in \RuNNer are multiplied by a cutoff function $f_{\mathrm{cut}}(r_{ij})$, which smoothly decays to zero at the cutoff radius, $f_{\mathrm{cut}}(r_{ij} \ge r_{\mathrm{cut}}) = 0$.
An optional inner cutoff radius $f(r_{ij} \le r_\mathrm{inner}) = 1$ shifts the onset of the decay away from the atom, which can be useful for feature sets targeting a specific distance range and for ignoring very short interatomic distances that are not present in chemically meaningful geometries.
When $r_\mathrm{inner} = 0$ (the default), the expressions below simplify to their standard single-parameter forms.

\paragraph{Cosine cutoff.}
\label{section:si:features:cutoff:cosine}
The cosine cutoff~\cite{P2882} provides a smooth, infinitely differentiable transition,
\begin{equation}
  f_{\mathrm{cut}}^\mathrm{cos}(r_{ij}) = \frac{1}{2}
    \left[\cos\!\left(\frac{\pi(r_{ij} - r_\mathrm{inner})}{r_{\mathrm{cut}} - r_\mathrm{inner}}\right) + 1\right], \quad r_\mathrm{inner} \le r_{ij} \le r_{\mathrm{cut}} \,.
\end{equation}

\paragraph{Hyperbolic tangent cutoff.}
\label{section:si:features:cutoff:tanh}
An alternative smooth cutoff based on the hyperbolic tangent is defined as
\begin{equation}
  f_{\mathrm{cut}}^\mathrm{tanh}(r_{ij}) =
    \tanh^3\!\left(1 - \frac{r_{ij} - r_\mathrm{inner}}{r_{\mathrm{cut}} - r_\mathrm{inner}}\right) \,,
    \quad r_\mathrm{inner} \le r_{ij} \le r_{\mathrm{cut}} \,,
\end{equation}
which decays smoothly to zero at $r_{\mathrm{cut}}$; the cube ensures continuous first and second derivatives at the cutoff, and at the inner radius the function takes the value $\tanh^3(1) \approx 0.50$.
Because the exact $\tanh$ function cannot be auto-vectorized by many compilers, \RuNNer also provides an approximation to this function that can be computed efficiently,
\begin{equation}
  \tanh(x) \approx \frac{x(27 + x^2)}{27 + 9x^2} \,,
\end{equation}
and enables SIMD vectorization of inner descriptor loops.

\paragraph{Polynomial cutoff.}
\label{section:si:features:cutoff:poly}
The polynomial cutoff
\begin{equation}
  f_{\mathrm{cut}}^\mathrm{poly}(r_{ij}) = \left(1 - \frac{r_{ij}^2}{r_{\mathrm{cut}}^2}\right)^n
\end{equation}
with integer exponent $n$ is used exclusively by the overlap matrix descriptors described in Sec.~\ref{section:si:features:overlap_matrix}.
It decays algebraically rather than exponentially and is differentiable up to order $n$ at $r_{\mathrm{cut}}$.

\paragraph{van der Waals cutoff.}
\label{section:si:features:cutoff:vdw}
For van der Waals interactions, a four-region smooth step function is required that is flat both near $r = 0$ and beyond $r_{\mathrm{cut}}$~\cite{PhysRevB.104.054106}.
The transition from zero to one begins at an inner radius $r_\mathrm{inner}$ extending over a buffer region of width $d_\mathrm{inner}$, and the following transition from one to zero begins at $r_{\mathrm{cut}} - d_\mathrm{out}$ over a region of width $d_\mathrm{out}$,
\begin{equation}
  f_{\mathrm{cut}}^\mathrm{vdW}(r) =
  \begin{cases}
    0 & r_{ij} \le r_\mathrm{inner} \\
    3t_\mathrm{inner}^2 - 2t_\mathrm{inner}^3 & r_\mathrm{inner} < r_{ij} \le r_\mathrm{inner} + d_\mathrm{inner} \\
    1 & r_\mathrm{inner} + d_\mathrm{inner} < r_{ij} \le r_{\mathrm{cut}} - d_\mathrm{out} \\
    1 - 3t_\mathrm{out}^2 + 2t_\mathrm{out}^3 & r_{\mathrm{cut}} - d_\mathrm{out} < r_{ij} \le r_{\mathrm{cut}} \\
    0 & r_{ij} > r_{\mathrm{cut}}
  \end{cases}
\end{equation}
with $t_\mathrm{inner} = (r_{ij} - r_\mathrm{inner})/d_\mathrm{inner}$ and $t_\mathrm{out} = (r_{ij} - r_{\mathrm{cut}} + d_\mathrm{out})/d_\mathrm{out}$.

\subsubsection{Atom-Centered Symmetry Functions}
\label{section:si:features:acsf}

Atom-centered symmetry functions (ACSFs)~\cite{P2882} map the local environment of atom $i$ onto a vector of scalars that are invariant under rotation, translation, and permutation of same-species neighbors~\cite{P2882}.
Each component $G_\mu$ of the feature vector depends on one or more hyperparameters controlling the spatial resolution and angular sensitivity of the corresponding feature.
\RuNNer implements five ACSF types covering radial and angular correlations. We follow the nomenclature in~\citenum{P2882}.

\paragraph{Radial atom-centered symmetry functions.}
\label{section:si:features:acsf:radial}
Type 2 ACSFs are Gaussians centered at a distance $R_{\mathrm{shift}}$ and broadened by a width parameter $\eta$,
\begin{equation}
  G_{i,\mu}^{(2)} = \sum_{j \in \mathcal{N}_i}
    \exp\!\left[-\eta_\mu \left(r_{ij} - r_{s,\mu}\right)^2\right] f_{\mathrm{cut}}(r_{ij}) \,.
  \label{eq:g2:si}
\end{equation}
They include contributions from all neighbors $\mathcal{N}_i$. By choosing a range of $(\eta_\mu, R_{\mathrm{shift},\mu})$ pairs, the radial distribution can be simultaneously resolved at multiple scales.

Type 1 ACSFs, a modified version of those described in~\citenum{P2882}, form a different radial basis by taking successive even powers of the cutoff function evaluated at $r_{ij}$,
\begin{equation}
  G_{i,\mu}^{(1)} = \sum_{j \neq i} \left[f_{\mathrm{cut}}(r_{ij})\right]^{2^{\mu - 1}} \,.
\end{equation}
The features are computed recursively as $G_{i,\mu}^{(1)} = \sum_{j \neq i}\bigl[G_{ij,\mu-1}^{(1)}\bigr]^2$ (starting from $G_1^{(1)} = \sum_{j\neq i} f_{\mathrm{cut}}(r_{ij})$), which makes the evaluation efficient while generating a set of basis functions with progressively shifting sensitivity from longer to shorter interatomic distances.

\paragraph{Angular atom-centered symmetry functions.}
\label{section:si:features:acsf:angular}
Type 3 ACSFs encode three-body distributions through the angle $\theta_{ijk}$ centered at atom $i$ and formed by neighbors $j$ and $k$~\cite{P2882},
\begin{equation}
  G_{i,\mu}^{(3)} = 2^{1-\zeta_\mu}
    \sum_{\substack{j,k \neq i \\ j < k}}
    \left(1 + \lambda_\mu \cos\theta_{ijk}\right)^{\zeta_\mu}
    \exp\!\left[-\eta_\mu\left(r_{ij}^2 + r_{ik}^2 + r_{jk}^2\right)\right]
    f_{\mathrm{cut}}(r_{ij})\, f_{\mathrm{cut}}(r_{ik})\, f_{\mathrm{cut}}(r_{jk}) \,,
  \label{eq:g4}
\end{equation}
where $\lambda_\mu \in \{-1, +1\}$ selects the parity of the angular term and $\zeta_\mu > 0$ sharpens the angular resolution.
The prefactor $2^{1-\zeta_\mu}$ normalizes the feature to approach unity at $\theta = 0$ for $\lambda = 1$.

Type 9 ACSFs~\cite{P2882} simplify type 3 by dropping the $r_{jk}$ distance dependence,
\begin{equation}
  G_{i,\mu}^{(9)} = 2^{1-\zeta_\mu}
    \sum_{\substack{j,k \neq i \\ j < k}}
    \left(1 + \lambda_\mu \cos\theta_{ijk}\right)^{\zeta_\mu}
    \exp\!\left[-\eta_\mu\left(r_{ij}^2 + r_{ik}^2\right)\right]
    f_{\mathrm{cut}}(r_{ij})\, f_{\mathrm{cut}}(r_{ik}) \,.
  \label{eq:g5}
\end{equation}
Because the third distance $r_{jk}$ neither enters the Gaussian envelope nor the cutoff product, type-9 functions are computationally cheaper while still encoding angular information through $\cos\theta_{ijk} = (\mathbf{r}_{ij} \cdot \mathbf{r}_{ik})/(r_{ij}\, r_{ik})$. Moreover, by omitting the third interatomic distance, more information is included as triples with large $r_{jk}$ are included.

Type 8 ACSFs place explicit Gaussians in angle space rather than in $\cos\theta$ space,
\begin{align}
  G_{i,\mu}^{(8)} = \sum_{\substack{j,k \neq i \\ j < k}}
    \Bigl[
      &e^{-\eta_\mu(\theta_{ijk} - \theta_{s,\mu})^2} +
       e^{-\eta_\mu(\theta_{ijk} - 360^\circ + \theta_{s,\mu})^2} \notag \\
      +\,&e^{-\eta_\mu(\theta_{ijk} + \theta_{s,\mu})^2} +
       e^{-\eta_\mu(\theta_{ijk} - 360^\circ - \theta_{s,\mu})^2}
    \Bigr]
    f_{\mathrm{cut}}(r_{ij})\, f_{\mathrm{cut}}(r_{ik})\, f_{\mathrm{cut}}(r_{jk}) \,,
  \label{eq:g8}
\end{align}
where the four terms ensure periodicity at $0^\circ$ and $360^\circ$ and guarantee that the function is symmetric with respect to 180°.

For multi-component systems, each ACSF can be evaluated separately for each neighbor species, providing element-resolved radial and angular channels.
This conventional construction causes the number of radial channels to grow linearly and the number of angular channels to grow approximately quadratically with the number of chemical species. \RuNNer therefore also provides element-embracing prefactors~\cite{eckhoff_lifelong_2023} for all ACSFs, which decouple the size of the feature vector from the number of chemical species in the dataset (see Sec.\;\ref{section:si:features:modifiers:ee}).

\subsubsection{Overlap Matrix Descriptors}
\label{section:si:features:overlap_matrix}

Overlap matrix descriptors~\cite{sadeghiMetricsMeasuringDistances2013,parsaeifardAssessmentStructuralResolution2021} encode the local chemical environment of atom $i$ through the eigenvalue spectrum of a matrix of atomic orbital overlap integrals, providing a continuously differentiable, rotationally invariant feature set using a typical basis in quantum chemistry.
Element information enters through the element-dependent basis function parameters assigned to each atom, so mixed-element environments are represented in one common matrix rather than through separate feature blocks depending on element combinations. 

\paragraph{Construction.}
\label{section:si:features:overlap_matrix:construction}
A set of Gaussian-type orbital (GTO) basis functions is attached to each neighbor of atom $i$ within the cutoff sphere.
The overlap matrix elements $S_{j\nu, k\mu}$ are given by the integral of the product of two GTOs $\nu$ and $\mu$ centered on atoms $j$ and $k$, respectively,
\begin{equation}
  S_{j\nu, k\mu} = \langle \phi_{j\nu} | \phi_{k\mu} \rangle \,.
  \label{eq:om_overlap}
\end{equation}
For $s$-type orbitals with exponents $\alpha_j$ and $\alpha_k$, the analytic overlap is
\begin{equation}
  \langle \phi^s_j | \phi^s_k \rangle =
    \left(\frac{\pi}{\alpha_j + \alpha_k}\right)^{3/2}
    \exp\!\left(-\frac{\alpha_j \alpha_k}{\alpha_j + \alpha_k}\, r_{jk}^2\right) \,.
\end{equation}
Mixed $s$--$p$ and $p$--$p$ overlaps introduce additional factors linear and bilinear in the displacement vector $\mathbf{r}_{jk}$, respectively, and are evaluated analytically.
The polynomial cutoff $f_{\mathrm{cut}}^\mathrm{poly}$ with its algebraic decay is employed here because it vanishes smoothly along with the orbital overlap as $r_{jk} \to r_{\mathrm{cut}}$.

\paragraph{Eigenvalue truncation (original OM feature).}
\label{section:si:features:overlap_matrix:trunc1}
After diagonalizing $\widetilde{S}^i \mathbf{U}^i = \mathbf{U}^i \boldsymbol{\Lambda}^i$, the feature values are taken as the $n_\mathrm{\lambda}$ largest eigenvalues $\lambda^i_1 \ge \lambda^i_2 \ge \cdots$; all remaining eigenvalues (corresponding to atoms that have not entered the cutoff sphere) are padded with zero.
This ensures that the feature vector length is independent of the number of neighbors.

\paragraph{Smooth variant.}
\label{section:si:features:overlap_matrix:trunc2}
The smooth variant maps the eigenvalue spectrum onto a fixed grid via a discrete sine transform, avoiding the discontinuity that occurs in the original variant when the eigenvalue count changes as atoms cross the cutoff boundary, see Eq.\;10 in the main text.

\paragraph{Derivatives.}
\label{section:si:features:overlap_matrix:derivatives}
Forces and stress require the derivatives of the features with respect to atomic positions.
For the smooth variant, differentiating Eq.~10 in the main text through the eigenvalue dependence yields
\begin{equation}
  \frac{\partial \widetilde{g}^i_{l^{\prime}}}{\partial \mathbf{r}_k}
  = \sum_{l=1}^{n_\lambda}
    \cos\!\left(\lambda^i_l \cdot x_{l^{\prime}}\right)\, x_{l^{\prime}}\;
    \frac{\partial \lambda^i_l}{\partial \mathbf{r}_k} \,,
  \label{eq:om_gridmap_deriv_runner}
\end{equation}
where $\mathbf{r}_k$ denotes a Cartesian coordinate.
The eigenvalue derivative follows from the standard identity for symmetric matrices,
\begin{equation}
  \mathrm{d}\lambda^i_l
  =
  \left(\mathbf{U}^i_l\right)^{\!T}
  \left(\mathrm{d}\widetilde{S}^i\right)
  \mathbf{U}^i_l \,,
  \label{eq:om_dlambda_runner}
\end{equation}
with $\mathbf{U}^i_l$ being the $l$-th eigenvector.
The differential $\mathrm{d}\widetilde{S}^i$ is obtained from Eq.~\ref{eq:om_overlap} by applying the product rule, combining derivatives of the GTO integrals with respect to $\mathbf{r}_{k}$ and the derivative of the cutoff function.

\subsubsection{Feature Modifiers}
\label{section:si:features:modifiers}

The additive terms of any ACSF can be multiplied in an element-specific way by a scalar prefactor that encodes additional physical information not captured by interatomic distances and angles alone.

\paragraph{Spin prefactor.}
\label{section:si:features:modifiers:spin}
For magnetic systems, the local spin moment $s_i$ of each atom is discretized to $s_i \in \{-1, 0, +1\}$ by applying a threshold whose magnitude depends on the training dataset: atoms whose magnetic moment $|m_i|$ lies below the threshold are treated as non-magnetic ($s_i = 0$), while all others carry $s_i = \mathrm{sign}(m_i)$.
For radial features involving a central atom $i$ and one neighbor $j$, three spin channels are defined~\cite{P6057},
\begin{align}
  M^+ &= \tfrac{1}{2}\,|s_i s_j|\,|s_i + s_j| \,, \\
  M^- &= \tfrac{1}{2}\,|s_i s_j|\,|s_i - s_j| \,, \\
  M^{0*} &= 1 - |s_i s_j| \,.
\end{align}
$M^+$ selects ferromagnetically aligned pairs, $M^-$ selects antiferromagnetically aligned pairs, and $M^{0*}$ captures all pairs where at least one atom is non-magnetic.
For angular features involving two neighbors $j$ and $k$, higher-order spin channels $M^{++}$, $M^{--}$, $M^{+-}$, $M^{00*}$, $M^{2++}$, $M^{3++}$, $M^{2--}$, and $M^{3--}$ encode all spin combinations of the three-body cluster $(i, j, k)$ in an analogous way.

\paragraph{Element-embracing prefactor.}
\label{section:si:features:modifiers:ee}
Conventional ACSFs require a separate feature for each neighbor element pair (or triplet), causing the descriptor dimensionality to scale with the number of element combinations.
Element-embracing ACSFs (eeACSFs)~\cite{eckhoff_lifelong_2023} eliminate this overhead by multiplying each pairwise contribution by an element-derived weight $H_{i,j}^\mathrm{rad}$ instead of partitioning the sum over element species.
For radial eeACSF type 2, \RuNNer employs
\begin{equation}
G_{i,\mu}^\mathrm{(2)} = \frac{1}{H_{\max,i}^\mathrm{rad}} \sum_{j \neq i} H_{i,j}^\mathrm{rad}
    \exp\!\left[-\eta_\mu \left(r_{ij} - r_{s,\mu}\right)^2\right] f_{\mathrm{cut}}(r_{ij})\;,
  \label{eq:eeacsf_rad}
\end{equation}
where $H_{\max,i}^\mathrm{rad}$ is the maximum possible value of $H_{i,j}^\mathrm{rad}$ for the chosen prefactor type, keeping all contributions in the range $[0,1]$ before summation.
An analogous formula applies to angular eeACSFs; see Eqs.~(4) and (7)--(9) of Ref.~\citenum{eckhoff_lifelong_2023}.

The weight $H_{i,j}^\mathrm{rad}$ is chosen from the set of element descriptors derived from the periodic table~\cite{eckhoff_lifelong_2023},
\begin{equation}
  H_{i,j}^\mathrm{rad} \in
  \bigl\{
    1,\;
    n_j,\;
    m_j,\;
    d_j,\;
    \bar{n}_j := X - n_j,\;
    \bar{m}_j := 9 - m_j,\;
    \bar{d}_j := 11 - d_j
  \bigr\} ,
  \label{eq:eeacsf_weights}
\end{equation}
where $n_j$ is the period of atom $j$, $m_j$ its position in the combined s/p-block (1 for group~1 through 8 for group~18, with helium assigned $m=8$ rather than its formal group number), $d_j$ its position in the d-block (1--10 for groups~3--12, zero for all other elements), and $X$ is a user-supplied upper bound on the period (maximum period $+ 1$).
The complement descriptors $\bar{n}_j$, $\bar{m}_j$, and $\bar{d}_j$ mirror the corresponding direct descriptors in a reversed scale, balancing the contributions of light and heavy elements.
By the convention of Ref.~\citenum{eckhoff_lifelong_2023}, $\bar{m}_j = 0$ for d-block elements (i.e.\ the complement is zero where the direct value $m_j$ is already zero) and $\bar{d}_j = 0$ for main-group elements (where $d_j = 0$), so that each complement descriptor is active only for the block it complements.

\RuNNer extends the scheme of Ref.~\citenum{eckhoff_lifelong_2023} with two additional f-block weight types.
The f-block position $f_j$ runs from 1 to 14 (lanthanides and actinides), and its complement $\bar{f}_j := 15 - f_j$.
By the same zero convention, $\bar{f}_j = 0$ for non-f-block elements, and $\bar{m}_j = \bar{d}_j = 0$ for f-block elements.
With these additions \RuNNer supports all 118 elements.

\subsubsection{Feature Scaling}
\label{section:si:features:scaling}

Before training, each feature vector component $G_{i,\mu}$ is linearly mapped to a standard range of values based on the training set,
\begin{equation}
  \widetilde{G}_{i,\mu} =
    \frac{G_{i,\mu} - \bar{G}_\mu}{G_\mu^\mathrm{max} - G_\mu^\mathrm{min}} \,,
  \label{eq:feature_scaling}
\end{equation}
where $\bar{G}_\mu$, $G_\mu^\mathrm{max}$, and $G_\mu^\mathrm{min}$ are the dataset mean, maximum, and minimum of feature $\mu$ across all atoms of the relevant species.
The scaling factors are written to a \texttt{scaling.data} file and reloaded unchanged during prediction.

\subsection{Neural Network Architecture}
\label{section:si:nn}

\subsubsection{Feed-Forward Neural Networks}
\label{section:si:nn:ffnns}

Each atomic sub-network in \RuNNer is a fully connected, feed-forward neural network with $L$ layers.
Given the scaled descriptor vector $\widetilde{\mathbf{G}}_i$ as input, the output of layer $l$ is
\begin{align}
  \mathbf{z}^{(l)} &= W^{(l)}\, \mathbf{a}^{(l-1)} + \mathbf{b}^{(l)}, \\
  \mathbf{a}^{(l)} &= \sigma^{(l)}\!\left(\mathbf{z}^{(l)}\right),
  \label{eq:nn_forward}
\end{align}
with weight matrix $W^{(l)} \in \mathbb{R}^{n_l \times n_{l-1}}$, bias vector $\mathbf{b}^{(l)} \in \mathbb{R}^{n_l}$, and activation function $\sigma^{(l)}$ applied element-wise.
The input layer sets $\mathbf{a}^{(0)} = \widetilde{\mathbf{G}}_i$.
All nodes within one layer share the same activation function by default, but \RuNNer also supports assigning a distinct activation per node, which is commonly used for pre-activating descriptors before the first hidden layer or controlling the range of individual output nodes.

Forces are obtained analytically via the chain rule through all layers.
The Jacobian $\partial E_i / \partial \widetilde{\mathbf{G}}_i$ is computed in a single backward pass, and the force contribution on atom $k$ from atom $i$ follows as
\begin{equation}
  F_{k,\alpha}^{(i)} =
    -\sum_\mu \frac{\partial E_i}{\partial \widetilde{G}_{i,\mu}}
    \frac{\partial \widetilde{G}_{i,\mu}}{\partial R_{k,\alpha}} \,,
\end{equation}
where the feature derivatives $\partial \widetilde{G}_{i,\mu}/\partial R_{k,\alpha}$ are available analytically for all implemented feature types.

\subsubsection{Activation Functions}
\label{section:si:nn:act}

\RuNNer provides a series of activation functions.

\paragraph{Hyperbolic tangent.}
\label{section:si:nn:act:tanh}
$\sigma(x) = \tanh(x)$ is the default activation.
It maps the input to $(-1, 1)$, is smooth everywhere, and saturates for $|x| \gtrsim 2.65$.
The \textit{scaled hyperbolic tangent} $\sigma(x) = 1.59223\,\tanh(x)$ uses the same functional form but extends the output range to approximately $(-1.59, 1.59)$. See the SI of \citenum{eckhoff_lifelong_2023} for more details.

\paragraph{Softplus.}
\label{section:si:nn:act:softplus}
The function $\sigma(x) = \ln(1 + e^x)$ is a smooth approximation to the rectified linear unit.
It is evaluated in the numerically stable form
\begin{equation}
  \sigma(x) = \ln\!\left(1 + e^{-|x|}\right) + \max(x,0)
\end{equation}
to avoid overflow in the exponential.

\paragraph{ReLU.}
\label{section:si:nn:act:relu}
The Rectified Linear Unit (ReLU) is $\sigma(x) = \max(0, x)$, which is computationally cheap but has a hard saturation at $x < 0$.

\paragraph{Sigmoid.}
\label{section:si:nn:act:sigmoid}
This function is defined as $\sigma(x) = 1/(1 + e^{-x})$, evaluated in the equivalent stable form $\sigma(x) = e^x/(e^x + 1)$ for $x < 0$.

\paragraph{Square and Linear.}
\label{section:si:nn:act:squarelinear}
The activations $\sigma(x) = x^2$ and the $\sigma(x) = x$ complete the set. The former is typically used to constrain the hardness in 4G training to positive values, while the latter is commonly used in the output layer to produce an unbounded scalar.

\subsubsection{Weight Initialization}
\label{section:si:nn:init}

Suitable initial weights are important for avoiding saturated activations at the start of training.
\RuNNer supports three initialization strategies.

\paragraph{Nguyen-Widrow.}
\label{section:si:nn:act:init:nguyen_widrow}
The Nguyen--Widrow method~\cite{NguyenWidrow1990} distributes the weight vectors of each hidden neuron such that the corresponding activation functions cover the scaled input range approximately uniformly.
For a layer with $n_\mathrm{inner}$ inputs and $n_\mathrm{out}$ outputs, the magnitude of each weight vector is set to $\beta = 0.7\, n_\mathrm{out}^{1/n_\mathrm{inner}}$, and the weights are then normalized to that length.

\paragraph{Xavier.}
\label{section:si:nn:act:init:xavier}
Xavier initialization~\cite{Glorot2010} samples weights from a uniform distribution $W_{ij} \sim \mathcal{U}(-a, a)$ with
\begin{equation}
  a = \sqrt{\frac{6}{n_\mathrm{inner} + n_\mathrm{out}}} \,,
\end{equation}
which preserves the variance of activations and gradients across layers in networks with linear or tanh-like activations.
The Eckhoff variant~\cite{Eckhoff2021} adapts this formula to account for the specific feature normalization used in HDNNPs.

\subsection{Energies, Forces, and Stress}
\label{section:si:energy}

\subsubsection{Atomic Energy Decomposition}
\label{section:si:energy:atomic}

In 2G-HDNNPs, the total energy predicted by \RuNNer is the sum of atomic contributions.
During training, possible zeroth order reference values for the atomic energies $\varepsilon_0$, e.g. the energies of free atoms, may be subtracted from the target total energy for configuration $n$,
\begin{equation}
  \tilde{E}_\mathrm{tot}^{(n)} = E_\mathrm{tot}^{(n)} - \sum_{i=1}^{N_n} \varepsilon_{0, i} \,.
\end{equation}
The network is trained on $\tilde{E}_\mathrm{tot}^{(n)}/N^{(n)}$, i.e. per-atom energies, which keeps the magnitude of the loss function independent of system size.
Predictions and references are brought to the same units by a user-specified normalization factor before computing the loss.

\subsubsection{Atomic Forces}
\label{section:si:energy:forces}

Atomic forces are obtained as the analytic gradient of the total energy with respect to atomic positions,
\begin{equation}
  \mathbf{F}_j = -\frac{\partial E}{\partial \mathbf{r}_j}
  = -\sum_{i}^N\sum_\mu^{N_\mu}
    \frac{\partial E_i}{\partial \widetilde{G}_{i,\mu}}
    \frac{\partial \widetilde{G}_{i,\mu}}{\partial \mathbf{r}_j} \,.
  \label{eq:forces}
\end{equation}
The outer sum extends over all atoms $i$ whose descriptor depends on the position $\mathbf{r}_j$.

\subsubsection{Stress Tensor}
\label{section:si:energy:stress}

We define the stress tensor $\boldsymbol{\sigma}$ such that its elements are given by
\begin{equation}
    \sigma_{\mu \nu} = \frac{1}{\Omega} \left.\frac{\partial E}{\partial \epsilon_{\mu \nu}}\right|_{\epsilon = 0},
\end{equation}
where $\Omega$ is the cell volume and $\boldsymbol{\epsilon}$ is the strain tensor.
Hence, the derivative of the energy with respect to the lattice vectors is
\begin{equation}
  \frac{\mathrm{d}E}{\mathrm{d}A} = \Omega\, \boldsymbol{\sigma}\, A^{-T},
\end{equation}
where $A^{-T}$ is the inverse transpose of the matrix of lattice vectors.
$\boldsymbol{\sigma}$ is positive when the cell is under tension and negative under compression, in agreement with the convention used in materials science and many \textit{ab initio} codes, but opposite to the convention of RuNNer~1.
In the interface to LAMMPS~\cite{P4473} the virial $-\Omega\,\boldsymbol{\sigma}$ is required.

\subsection{Electrostatics}
\label{section:si:elec}

\subsubsection{Third-Generation HDNNPs}
\label{section:si:elec:3G}

In 3G-HDNNPs, each atom carries an environment-dependent partial charge $q_i$ predicted by an atomic neural network.
The total energy separates into a short-range part and a Coulomb part,
\begin{equation}
  E = E_\mathrm{short}(\{\widetilde{\mathbf{G}}_i\}) + E_\mathrm{elec}(\{q_i\}, \{\mathbf{R}_i\}) \,.
\end{equation}
Since both contributions depend on the atomic positions, and since in particular the charges are environment-dependent, the force on atom $j$ contains a direct term and a charge-mediated term,
\begin{equation}
  \mathbf{F}_j = -\frac{\partial E_\mathrm{short}}{\partial \mathbf{r}_j}
  - \left.
    \frac{\partial E_\mathrm{elec}}{\partial \mathbf{r}_j}\right|_{q}
  - \sum_i \frac{\partial E_\mathrm{elec}}{\partial q_i}
    \frac{\partial q_i}{\partial \mathbf{r}_j} \,,
  \label{eq:3g_forces}
\end{equation}
where the last term propagates the gradient through the charge network via its descriptor derivatives.
The constraint $\sum_i q_i = q_\mathrm{tot}$ is enforced during prediction by projecting the predicted charges onto the constraint hyperplane.

\subsubsection{Fourth-Generation HDNNPs and Charge Equilibration}
\label{section:si:elec:4G}

In 4G-HDNNP~\cite{P5932}, the charge distribution is determined via charge equilibration by minimizing the electrostatic energy with respect to the atomic charges subject to total charge conservation.
Two species-specific networks predict the atomic electronegativity $\chi_i$ and the hardness $J_i$, respectively.
The QEq electrostatic energy is given by
\begin{equation}
  E_\mathrm{QEq} = \sum_i \left(\chi_i q_i + \tfrac{1}{2} J_i q_i^2\right) + E_\mathrm{elec}(\{q_i\}, \{\mathbf{R}_i\}) \,,
\end{equation}
and the equilibrium charge vector $\mathbf{q} = (q_1, \ldots, q_N)$ satisfies
\begin{equation}
  (\mathbf{A} + \mathbf{J})\,\mathbf{q} = -\boldsymbol{\chi} + \mu\,\mathbf{1} \,,
  \label{eq:qeq}
\end{equation}
where $\mathbf{A}$ is the Coulomb matrix with elements $A_{ij} = \partial^2 E_\mathrm{elec}/\partial q_i \partial q_j$, $\mathbf{J} = \mathrm{diag}(J_1, \ldots, J_N)$, and $\mu$ is a Lagrange multiplier enforcing $\sum_i q_i = q_\mathrm{tot}$.

The force on atom $k$ carries contributions from the position dependence of the Coulomb matrix $\mathbf{A}(\{\mathbf{R}_i\})$, the electronegativities $\chi_i(\mathbf{R}_k)$, and the hardnesses $J_i(\mathbf{R}_k)$.
These are computed via the ``force trick'': the Lagrange multipliers $\boldsymbol{\lambda}$ satisfying $(\mathbf{A} + \mathbf{J})^T \boldsymbol{\lambda} = -\partial E/\partial \mathbf{q}$ are solved once per structure, and the total force is then assembled as
\begin{equation}
  \mathbf{F}_k = -\frac{\partial E_\mathrm{short}}{\partial \mathbf{R}_k}
    - \left.\frac{\partial E_\mathrm{elec}}{\partial \mathbf{R}_k}\right|_{q}
    - \boldsymbol{\lambda}^T \frac{\partial \mathbf{A}}{\partial \mathbf{R}_k}\,\mathbf{q}
    - \sum_i \left(\frac{\partial E}{\partial q_i} + \lambda_i\right)
      \frac{\partial (J_i q_i + \chi_i)}{\partial \mathbf{R}_k} \,.
  \label{eq:4g_force_trick}
\end{equation}
This avoids the need to differentiate through the QEq linear solver, reducing the computational cost to a single additional system of linear equations of the same size as in Eq.~\ref{eq:qeq}.

\subsubsection{Ewald Summation}
\label{section:si:elec:ewald}

For periodic systems, the Coulomb energy of a charge distribution $\{q_i\}$ with Gaussian charge widths $\sigma_i$ is evaluated using Ewald summation~\cite{ewald_berechnung_1921}.
The total interaction is split into a real-space sum that converges rapidly in direct space and a reciprocal-space sum that converges rapidly in Fourier space,
\begin{equation}
  E_\mathrm{elec} = E_\mathrm{real} + E_\mathrm{recip} + E_\mathrm{self} \,,
\end{equation}
where the splitting parameter $\eta$ controls the balance between the two contributions and is chosen automatically to equalize the cost of the two sums at a given target precision.
The real-space sum decays as $\mathrm{erfc}(\eta r)/r$, while the reciprocal-space sum is evaluated as a single Fourier transform of the charge density times a Gaussian filter factor $\exp(-G^2/(4\eta^2))$.
Gaussian charges with width $\sigma_i$ are accommodated by replacing the point-charge erfc kernel with the corresponding Gaussian convolution, which modifies the self-energy correction $E_{\mathrm{self}}$ and the short-range real-space term.

For the QEq charge equilibration, \RuNNer also computes the Coulomb influence matrix $\mathbf{A}$ and its derivatives with respect to atomic positions analytically from the same Ewald framework.

\subsubsection{Particle--Particle--Particle--Mesh Method}
\label{section:si:elec:pppm}

As an alternative to the classical Ewald summation, which may become computationally demanding for large systems, \RuNNer implements a particle--particle--particle--mesh (PPPM) solver~\cite{Hockney1988} that offloads the reciprocal-space contribution to a three-dimensional fast Fourier transform (FFT) on a regular grid.
The charge density is interpolated onto the grid, the Poisson equation is solved in Fourier space, and the result is interpolated back to the particle positions.
The splitting parameter defaults to $\eta = 2.2\,\mathrm{Bohr}^{-1}$, or is set to the maximum Gaussian width $\max_i \sigma_i$ when Gaussian charges are used.
The FFT grid size is chosen such that grid points per Gaussian standard deviation satisfies a precision criterion, and the grid dimensions are restricted to multiples of 4 for efficiency.
For the QEq charge equilibration, a separate FFT-based Lagrange solver is used to impose the total-charge constraint directly in Fourier space.

\subsection{Two-Body Repulsion}
\label{section:si:twobody}

\subsubsection{Ziegler--Biersack--Littmark Potential}
\label{section:si:twobody:zbl}

Short-range nuclear repulsion between atomic cores can be added to any HDNNP model through the Ziegler--Biersack--Littmark (ZBL) potential~\cite{ziegler_stopping_1985}.
The ZBL energy for a pair of atoms $i$ and $j$ separated by $r_{ij}$ is
\begin{equation}
  E_\mathrm{ZBL}^{ij} = \frac{Z_i Z_j}{r_{ij}}\,\Phi(x)\,f_{\mathrm{cut}}(r_{ij}) \,,
  \label{eq:zbl}
\end{equation}
where $Z_i$ and $Z_j$ are the atomic numbers of atoms $i$ and $j$, and atomic units ($e = 1$, energies in Hartree, distances in Bohr) are used throughout.
The dimensionless screening coordinate is
\begin{equation}
  x = \frac{Z_i^{0.23} + Z_j^{0.23}}{0.88533}\,r_{ij} \,,
\end{equation}
where the denominator is the Lindhard screening length in Bohr.
The universal screening function is a sum of four exponentials fitted to a large database of ion--atom scattering experiments,
\begin{equation}
  \Phi(x) = 0.18175\,e^{-3.19980\,x}
           + 0.50986\,e^{-0.94229\,x}
           + 0.28022\,e^{-0.40290\,x}
           + 0.02817\,e^{-0.20162\,x} \,.
  \label{eq:zbl_screening}
\end{equation}
The four coefficients in Eq.~\ref{eq:zbl_screening} sum to unity, so $\Phi(0) = 1$ and the potential reduces to the bare Coulomb interaction at the origin.
Forces are obtained analytically from the derivatives of Eqs.~\ref{eq:zbl}--\ref{eq:zbl_screening} with respect to $r_{ij}$.
\subsection{van der Waals Interactions}
\label{section:si:vdw}

\subsubsection{Hirshfeld-Based Dispersion}
\label{section:si:vdw:hirsh}

Van der Waals dispersion interactions extending beyond the local atomic environments are accounted for via the Tkatchenko--Scheffler scheme~\cite{P2121}, which scales free-atom $C_6$ coefficients by the environment-dependent Hirshfeld volume ratio~\cite{hirshfeld_bonded_1977}.
Each atom $a$ is assigned an effective volume $V_a$ predicted by a species-specific neural network, and the corresponding effective vdW radius and $C_6$ coefficient are
\begin{align}
  R_a^\mathrm{eff} &= V_a^{1/3}\, R_a^\mathrm{free} \,, \\
  C_6^{aa} &= V_a^2\, C_6^{aa,\mathrm{free}} \,.
\end{align}
The cross-species coefficient is obtained from the London combination rule,
\begin{equation}
  C_6^{ab} = \frac{2\,C_6^{aa,\mathrm{free}}\,C_6^{bb,\mathrm{free}}}
    {\dfrac{\alpha_b^\mathrm{free}}{\alpha_a^\mathrm{free}}\,C_6^{aa,\mathrm{free}} +
     \dfrac{\alpha_a^\mathrm{free}}{\alpha_b^\mathrm{free}}\,C_6^{bb,\mathrm{free}}}
  \cdot V_a\, V_b \,,
  \label{eq:c6_cross}
\end{equation}
where $\alpha_s^\mathrm{free}$ are the free-atom polarizabilities.
The dispersion energy for a pair $(a, b)$ is
\begin{equation}
  E_\mathrm{vdW}^{ab} = -C_6^{ab}\,f_\mathrm{damp}(r_{ab})\, r_{ab}^{-6}\, f_{\mathrm{cut}}(r_{ab}) \,,
\end{equation}
where the Fermi-type damping function
\begin{equation}
  f_\mathrm{damp}(r_{ab}) =
    \frac{1}{1 + \exp\!\left[-d\!\left(\dfrac{r_{ab}}{s_R\, R_{ab}^\mathrm{eff}} - 1\right)\right]}
\end{equation}
with $R_{ab}^\mathrm{eff} = R_a^\mathrm{eff} + R_b^\mathrm{eff}$ suppresses the divergence at short range.
The parameters $d$ and $s_R$ control the onset and steepness of the damping.
Forces include gradient contributions from the volume dependence of the effective radii and $C_6$ coefficients; the van der Waals cutoff function $f_{\mathrm{cut}}^\mathrm{vdW}$ (Sec.~\ref{section:si:features:cutoff:vdw}) is used for $f_{\mathrm{cut}}(r_{ab})$.

\subsection{Training}
\label{section:si:training}

\subsubsection{Loss Functions}
\label{section:si:training:loss}

\RuNNer supports three loss functions for measuring discrepancies between predicted and reference values.
Let $\hat{y}_s$ and $y_s$ denote the predicted and reference values for sample $s$, and let $S$ be the number of samples in the current batch.

The \textit{mean squared error} (MSE) uses a prefactor of $\tfrac{1}{2}$ so that its gradient equals the signed residual,
\begin{equation}
  \mathcal{L}_\mathrm{MSE} = \frac{1}{S}\sum_{s=1}^S \frac{1}{2}(\hat{y}_s - y_s)^2 \,, \quad
  \frac{\partial \mathcal{L}_\mathrm{MSE}}{\partial \hat{y}_s} = \frac{\hat{y}_s - y_s}{S} \,.
\end{equation}

The \textit{mean absolute error} (MAE) penalizes residuals linearly and is more robust to outliers,
\begin{equation}
  \mathcal{L}_\mathrm{MAE} = \frac{1}{S}\sum_{s=1}^S |\hat{y}_s - y_s| \,, \quad
  \frac{\partial \mathcal{L}_\mathrm{MAE}}{\partial \hat{y}_s} = \frac{\mathrm{sign}(\hat{y}_s - y_s)}{S} \,.
\end{equation}

The \textit{root mean squared error} (RMSE) penalizes outliers more strongly,
\begin{equation}
  \mathcal{L}_\mathrm{RMSE} = \sqrt{\frac{1}{S}\sum_{s=1}^S (\hat{y}_s - y_s)^2} \,, \quad
  \frac{\partial \mathcal{L}_\mathrm{RMSE}}{\partial \hat{y}_s}
    = \frac{\hat{y}_s - y_s}{S\,\mathcal{L}_\mathrm{RMSE}} \,.
\end{equation}

\subsubsection{Optimization Algorithms}
\label{section:si:training:opt}

\paragraph{Gradient Descent.}
\label{section:si:training:opt:gd}
The standard gradient descent optimizer updates parameters along the mean negative gradient,
\begin{equation}
  \boldsymbol{\theta}_{t+1} = \boldsymbol{\theta}_t - \alpha\,\bar{\mathbf{g}}_t \,,
\end{equation}
where $\alpha$ is the learning rate and $\bar{\mathbf{g}}_t = n_s^{-1}\sum_{s=1}^{n_s} \nabla \mathcal{L}_s$ is the gradient averaged over $n_s$ streams (MPI tasks).

\paragraph{SGD with Momentum.}
\label{section:si:training:opt:sgdmom}
Stochastic gradient descent with momentum accumulates velocity from previous gradients to dampen oscillations and accelerate convergence along persistent descent directions,
\begin{align}
  \mathbf{v}_{t} &= \gamma\,\mathbf{v}_{t-1} - \alpha\,\bar{\mathbf{g}}_t \,, \\
  \boldsymbol{\theta}_{t+1} &= \boldsymbol{\theta}_t + \mathbf{v}_t \,,
\end{align}
with momentum coefficient $\gamma$ (default 0.9) and learning rate $\alpha$ (default 0.001).

\paragraph{Adam.}
\label{section:si:training:opt:adam}
Adam~\cite{P5404} maintains exponential moving averages of the first and second moments of the gradient with bias correction,
\begin{align}
  \mathbf{m}_t &= \beta_1\,\mathbf{m}_{t-1} + (1-\beta_1)\,\bar{\mathbf{g}}_t \,, \\
  \mathbf{v}_t &= \beta_2\,\mathbf{v}_{t-1} + (1-\beta_2)\,\bar{\mathbf{g}}_t^2 \,, \\
  \hat{\mathbf{m}}_t &= \mathbf{m}_t / (1 - \beta_1^t) \,, \quad
  \hat{\mathbf{v}}_t = \mathbf{v}_t / (1 - \beta_2^t) \,, \\
  \boldsymbol{\theta}_{t+1} &= \boldsymbol{\theta}_t
    - \alpha\,\hat{\mathbf{m}}_t \oslash \left(\sqrt{\hat{\mathbf{v}}_t} + \varepsilon\right) \,,
\end{align}
where $\oslash$ denotes element-wise division.
Default hyperparameters are $\beta_1 = 0.9$, $\beta_2 = 0.999$, $\varepsilon = 10^{-8}$, and $\alpha = 0.001$.

\paragraph{Multistream Kalman Filter.}
\label{section:si:training:opt:kalman}
Using the Kalman filter~\cite{P1308} in its multistream variant~\cite{P3962,singraber_parallel_2019}, \RuNNer treats neural network training as a state estimation problem. The optimizer maintains a symmetric covariance matrix $\mathbf{P}$ of shape $\left(n_p \times n_p\right)$ representing parameter uncertainty, where $n_p$ corresponds to the number of parameters. During training, the optimizer iteratively refines the parameters $\boldsymbol{\theta}$ using measurement data in the form of the loss and loss gradients from each stream. At the beginning of each iteration $t$, a scaling matrix $\mathbf{A}\left(t\right)$ of shape $\left(n_s \times n_s\right)$, where $n_s$ is the number of input streams, is computed as
\begin{equation}
\mathbf{A}\left(t\right) = \left( \eta\left(t\right) \mathbf{I} + \mathbf{H}\left(t\right)^T \mathbf{P}\left(t\right) \mathbf{H}\left(t\right) \right)^{-1},
\end{equation}
with $\mathbf{H}\left(t\right)$ being the Jacobian (gradients) of the loss from each stream of shape $\left(n_p \times n_s\right)$, $\eta\left(t\right)$ being measurement noise (or learning rate), and $\mathbf{I}$ being an identity matrix of shape $\left(n_s \times n_s\right)$.
The Kalman gain $\mathbf{K}\left(t\right)$ of shape $\left(n_p \times n_s\right)$ is obtained as
\begin{equation}
\mathbf{K}\left(t\right) = \mathbf{P}\left(t\right) \mathbf{H}\left(t\right) \mathbf{A}\left(t\right).
\end{equation}
Using $\mathbf{K}\left(t\right)$ and the loss from each stream $\boldsymbol{\xi}\left(t\right)$, the parameters are updated as
\begin{equation}
\boldsymbol{\theta}\left(t+1\right) = \boldsymbol{\theta}\left(t\right) - \mathbf{K}\left(t\right) \cdot \boldsymbol{\xi}\left(t\right),
\end{equation}
and the updated covariance matrix is
\begin{equation}
\mathbf{P}\left(t+1\right) = \lambda^{-1} \left( \mathbf{P}\left(t\right) - \mathbf{K}\left(t\right) \mathbf{H}\left(t\right)^T \mathbf{P}\left(t\right) \right),
\end{equation}
where $\lambda$ is the forgetting factor.
If process noise is enabled, the covariance matrix is augmented by
\begin{equation}
\mathbf{P}\left(t\right) \leftarrow \mathbf{P}\left(t\right) + q\left(t\right) \mathbf{I},
\end{equation}
where $q\left(t\right)$ corresponds to the noise level.

\paragraph{Hyperparameter scheduling.}
\label{section:si:training:opt:scheduling}
Both Kalman variants schedule their hyperparameters during training. In the standard Kalman filter the measurement noise $\eta$ grows exponentially toward a ceiling value, and an optional process noise term $q$ decays exponentially—together these prevent the covariance matrix from collapsing prematurely and allow the filter to remain responsive to new data. In the fading-memory variant the forgetting factor $\lambda$ is updated each step according to
\begin{equation}
\lambda(t+1) = \lambda(t)\cdot\lambda_0 + (1-\lambda_0),
\end{equation}
which drives $\lambda$ toward unity, progressively reducing the weight given to old observations and keeping the covariance matrix from becoming singular in long training runs.

\paragraph{Regularization of the scaling matrix.}
\label{section:si:training:opt:regularization}
To prevent numerical instabilities when forming the scaling matrix $\mathbf{A}(t)$, \RuNNer spectrally regularizes the symmetric matrix $\mathbf{M}(t) = \eta(t)\,\mathbf{I} + \mathbf{H}(t)^T \mathbf{P}(t)\, \mathbf{H}(t)$ that is inverted to obtain $\mathbf{A}(t) = \mathbf{M}(t)^{-1}$. After eigendecomposition $\mathbf{M} = \mathbf{U}\mathbf{D}\mathbf{U}^T$, each eigenvalue $d_i$ is replaced by
\begin{equation}
\tilde{d}_i = d_i + \varepsilon\exp\!\left(-\frac{d_i}{\varepsilon}\right),
\end{equation}
which leaves large eigenvalues unchanged while lifting near-zero eigenvalues to $\varepsilon$, keeping the inverse well-conditioned. Closely spaced eigenvalues that could cause degenerate eigenvector problems receive an additional small random perturbation.

\paragraph{Model-Optimizer Mapping and Correlation Matrix Dimensionality}
In \RuNNer, optimizers can be flexibly assigned to one or multiple models. Furthermore, optimizers can be assigned to the individual atomic neural networks contained in one HDNN model. As a result, users can freely choose to update all elements with a single optimizer (global update step) or update them one by one (element-wise update step). In case of the Kalman filter optimizer, this influences the size of the correlation matrix: for a global update step, the off-diagonal terms correspond to cross-element couplings. We generally find that this has a strongly positive influence on training convergence.

\paragraph{Gradient clipping.}
\label{section:si:training:opt:clip}
To stabilize training when large parameter updates occur, \RuNNer supports two gradient clipping strategies.
\textit{Value clipping} replaces each component individually: $g_i \leftarrow \mathrm{clamp}(g_i, -c, c)$.
\textit{Norm clipping} rescales the entire gradient vector when its $\ell^2$ norm exceeds a threshold $c_\mathrm{norm}$,
\begin{equation}
  \mathbf{g} \leftarrow \mathbf{g} \cdot \frac{c_\mathrm{norm}}{\|\mathbf{g}\|_2}
  \quad \text{if} \quad \|\mathbf{g}\|_2 > c_\mathrm{norm} \,,
\end{equation}
which preserves the direction of the gradient while restricting its magnitude.

\subsection{Atomic Reference Data}
\label{section:si:periodic_table}

Several physical models in \RuNNer require element-specific reference data.
Covalent radii for all 118 elements are taken from the WebElements database~\cite{WebElements}, stored in Bohr.
Van der Waals radii, free-atom static polarizabilities $\alpha_s^\mathrm{free}$, and homonuclear $C_6$ coefficients $C_6^{ss,\mathrm{free}}$ enter the Hirshfeld dispersion model (Sec.~\ref{section:si:vdw:hirsh}) and are taken from Hermann \textit{et al.}~\cite{hermann_libmbd}. Pearson electronegativity and hardness values are used for initializing the respective models during 4G training and are taken from the 97th edition CRC Handbook of Chemistry and Physics~\cite{CRCHandbook97}.
All quantities are stored in atomic units internally and converted from input and for output if required.

\section{Methods}
\label{sec:si:methods}

\subsection{Dataset Generation and HDNNP Training}
\label{sec:si:methods:dataset}

The water dataset used for all performance benchmarks was drawn from the dataset of Thiemann \textit{et al.}~\cite{thiemann_water_2022}.
We retained only the pure liquid water configurations, yielding 140 structures of 96 atoms each.
All structures were recalculated at the RPBE level~\cite{hammer_rpbe} using FHI-aims~\cite{P2189} (version \texttt{221103}) with the \texttt{intermediate} numerical atom-centered orbital basis set.
Scalar-relativistic effects were included at the atomic ZORA level.
Electronic self-consistency was converged to $5 \times 10^{-6}\,e\,\mathrm{Bohr}^{-3}$, $10^{-3}\,\mathrm{eV}$, and $10^{-6}\,\mathrm{eV}$ in the electron density, sum of eigenvalues, and total energy, respectively.
The Brillouin zone was sampled with a $k$-point density of $4.0\,\mathrm{Bohr}^{-1}$ per reciprocal lattice vector using a Monkhorst--Pack grid.
Fermi--Dirac smearing with a width of $0.01\,\mathrm{eV}$ was applied.
The Kohn--Sham equations were solved with the ELPA eigensolver~\cite{elpa_marek_2014} using the LVL\_fast resolution-of-identity approximation and a Pulay mixer with a history of 12 steps and a mixing parameter of 0.2.
Hirshfeld volume ratios were computed at every SCF step (\texttt{hirshfeld\_always}) for use as 3G training targets.
The dataset was partitioned randomly into a training set of 126 structures and a test set of 14 structures.

For each of the three HDNNP generations (2G, 3G, 4G), a committee of eight independently initialized models was trained using \RuNNer (version \texttt{2.0.0-alpha.3}, Git commit \texttt{e3232d79}).
The training protocol for each generation was as follows.
For the 2G-HDNNP, a single round of energy and force training was performed.
For the 3G-HDNNP, the charge model was first trained on reference atomic charges from the electronic structure calculations, followed by a second stage in which the short-range energy network was trained on the residual energies and forces after subtracting the electrostatic contribution.
For the 4G-HDNNP, electronegativity and hardness networks were first trained on reference charges, followed by short-range energy and force training.
All training stages used the multistream Kalman filter optimizer.
The model at the epoch with the lowest RMSE on the test set was selected for benchmarking.
The resulting training and test RMSEs at the optimal epoch are reported in Tables~\ref{tab:2g_rmse}--\ref{tab:4g_rmse}.
Full training inputs and the dataset are available at \url{https:/gitlab.com/runner-suite/runner2-paper-data}.

\begin{table}[h]
  \centering
  \caption{%
    Training and test RMSEs of the eight 2G-HDNNP committee members at the optimal epoch (epoch 50).
    Energy RMSEs are per-atom; force RMSEs are per force component.\\
    \vspace{2pt}
    \small\textbf{Alt Text:} Training and test root mean squared errors for the eight 2G-HDNNP committee members trained on the water dataset for performance benchmarking.
  }
  \label{tab:2g_rmse}
  \begin{tabular}{cSSSS}
    \toprule
    & \multicolumn{2}{c}{Energy RMSE (meV\,atom$^{-1}$)}
    & \multicolumn{2}{c}{Force RMSE (meV\,\AA$^{-1}$)} \\
    \cmidrule(lr){2-3} \cmidrule(lr){4-5}
    Member & {Train} & {Test} & {Train} & {Test} \\
    \midrule
    1 & 0.0252 & 0.5415 & 67.90  & 76.16  \\
    2 & 0.0274 & 0.5227 & 67.51  & 77.43  \\
    3 & 0.0278 & 0.5544 & 66.33  & 75.40  \\
    4 & 0.0225 & 0.5077 & 65.56  & 72.28  \\
    5 & 0.0253 & 0.4959 & 65.72  & 73.24  \\
    6 & 0.0267 & 0.4683 & 66.09  & 73.48  \\
    7 & 0.0253 & 0.5772 & 69.10  & 75.69  \\
    8 & 0.0276 & 0.5437 & 69.00  & 76.89  \\
    \bottomrule
  \end{tabular}
\end{table}

\begin{table}[h]
  \centering
  \caption{%
    Training and test RMSEs of the eight 3G-HDNNP committee members at the optimal epoch.
    \textbf{(a)} Charge network at epoch 5.
    \textbf{(b)} Short-range energy network at epoch 49.
    Charge RMSEs are per atom; energy and force RMSEs are as in Table~\ref{tab:2g_rmse}.\\
    \vspace{2pt}
    \small\textbf{Alt Text:} Training and test root mean squared errors for the eight 3G-HDNNP committee members trained on the water dataset for performance benchmarking.
  }
  \label{tab:3g_rmse}
  \begin{tabular}{cSS}
    \toprule
    & \multicolumn{2}{c}{Charge RMSE ($m|e|$)} \\
    \cmidrule(lr){2-3}
    Member & {Train} & {Test} \\
    \midrule
    1 & 3.9214 & 5.0437 \\
    2 & 3.9298 & 4.9549 \\
    3 & 3.9020 & 4.9902 \\
    4 & 3.8976 & 4.9643 \\
    5 & 3.9368 & 5.0566 \\
    6 & 3.9273 & 4.9448 \\
    7 & 3.8941 & 4.9650 \\
    8 & 3.9225 & 4.9775 \\
    \midrule\midrule
    & \multicolumn{2}{c}{Energy RMSE (meV\,atom$^{-1}$)} \\
    \cmidrule(lr){2-3}
    Member & {Train} & {Test} \\
    \midrule
    1 & 1.0775 & 0.9304 \\
    2 & 1.0419 & 0.9086 \\
    3 & 1.0844 & 1.1062 \\
    4 & 1.1028 & 0.8745 \\
    5 & 1.0856 & 0.9024 \\
    6 & 1.0737 & 0.7919 \\
    7 & 1.0843 & 0.8573 \\
    8 & 1.0590 & 0.8561 \\
    \midrule\midrule
    & \multicolumn{2}{c}{Force RMSE (meV\,\AA$^{-1}$)} \\
    \cmidrule(lr){2-3}
    Member & {Train} & {Test} \\
    \midrule
    1 & 93.39  & 105.93 \\
    2 & 93.97  & 105.71 \\
    3 & 94.07  & 107.09 \\
    4 & 93.62  & 106.67 \\
    5 & 93.62  & 108.17 \\
    6 & 95.56  & 107.31 \\
    7 & 94.99  & 107.60 \\
    8 & 96.25  & 106.85 \\
    \bottomrule
  \end{tabular}
\end{table}

\begin{table}[h]
  \centering
  \caption{%
    Training and test RMSEs of the eight 4G-HDNNP committee members at the optimal epoch.
    \textbf{(a)} Charge network (electronegativity and hardness) at epoch 50.
    \textbf{(b)} Short-range energy network at epoch 27.
    Units are as in Table~\ref{tab:3g_rmse}.\\
    \vspace{2pt}
    \small\textbf{Alt Text:} Training and test root mean squared errors for the eight 4G-HDNNP committee members trained on the water dataset for performance benchmarking.
  }
  \label{tab:4g_rmse}
  \begin{tabular}{cSS}
    \toprule
    & \multicolumn{2}{c}{Charge RMSE ($m|e|$)} \\
    \cmidrule(lr){2-3}
    Member & {Train} & {Test} \\
    \midrule
    1 & 4.6233 & 5.3686 \\
    2 & 4.6308 & 5.2545 \\
    3 & 5.0193 & 5.5475 \\
    4 & 4.8361 & 5.5712 \\
    5 & 4.8925 & 5.5152 \\
    6 & 4.7631 & 5.3102 \\
    7 & 4.7806 & 5.4848 \\
    8 & 4.9763 & 5.5721 \\
    \midrule\midrule
    & \multicolumn{2}{c}{Energy RMSE (meV\,atom$^{-1}$)} \\
    \cmidrule(lr){2-3}
    Member & {Train} & {Test} \\
    \midrule
    1 & 1.2824 & 1.6233 \\
    2 & 1.3739 & 1.9319 \\
    3 & 1.3214 & 1.7160 \\
    4 & 1.2143 & 1.4117 \\
    5 & 1.3080 & 1.8020 \\
    6 & 1.2422 & 1.6382 \\
    7 & 1.2524 & 1.7528 \\
    8 & 1.2846 & 1.8885 \\
    \midrule\midrule
    & \multicolumn{2}{c}{Force RMSE (meV\,\AA$^{-1}$)} \\
    \cmidrule(lr){2-3}
    Member & {Train} & {Test} \\
    \midrule
    1 & 115.53 & 113.29 \\
    2 & 125.46 & 123.83 \\
    3 & 109.77 & 108.65 \\
    4 & 110.82 & 110.35 \\
    5 & 117.62 & 116.33 \\
    6 & 107.20 & 107.17 \\
    7 & 112.77 & 112.27 \\
    8 & 114.38 & 114.99 \\
    \bottomrule
  \end{tabular}
\end{table}

\subsection{OpenMP, MPI, and ASE Interface Benchmarks}
\label{sec:si:methods:perf_bechmarks}

The OpenMP thread-scaling and MPI strong-scaling benchmarks (Fig.~\ref{fig:lammps_performance}a and b) and the ASE interface committee-scaling benchmarks (Fig.~\ref{fig:ase_performance}) were all performed on the Otus cluster at the Paderborn Center for Parallel Computing (PC$^2$) using the same committee potential described in Sec.~\ref{sec:si:methods:dataset}.
Each node is equipped with two AMD EPYC 9655 processors providing 96 physical cores (192 hardware threads) at up to 4.5\,GHz and 768\,GiB of RAM.
Nodes are interconnected via Nvidia NDR InfiniBand at 200\,Gbps per port.

\RuNNer (version \texttt{2.0.0-alpha.11}, Git commit \texttt{ef43e5f0bf}) was compiled with GCC~14.3 (\texttt{mpicc} / \texttt{mpicxx} / \texttt{mpif90}), MKL~2025.2.0, and OpenMPI~5.0.8, all loaded from the cluster module system.
Compiler optimizations were set to \texttt{-O3} with link-time optimization enabled and \texttt{-march=native} to exploit the AVX-512 instruction set of the EPYC processors.
MKL was linked against the GNU threading layer (\texttt{MKL\_THREADING\_LAYER=GNU}) to enable consistent scaling of the internal FFTW routines under OpenMP.
The \RuNNer compile-time flag \texttt{FC\_NO\_INNER\_CUTOFF} was enabled to improve descriptor evaluation performance for the benchmark cutoff settings.

\paragraph{LAMMPS benchmarks (Fig.~\ref{fig:lammps_performance}a and b).}
\label{sec:si:methods:perf_bechmarks:lammps}
The simulation cell consisted of a single 96-atom bulk water unit cell replicated to 96\,000 atoms using ASE~\cite{P5915}.
All simulations used the NVT ensemble at 300\,K with full particle propagation driven by LAMMPS~\cite{P4473} (development version, git commit \texttt{0c44aae4cb}).
LAMMPS was built with the \texttt{OPT} and \texttt{OPENMP} packages; for the pure-OpenMP scaling tests the MPI-serial build of \RuNNer was used.
All benchmarks ran on exclusive nodes with \texttt{OMP\_PLACES=cores} and \texttt{OMP\_PROC\_BIND=close}; LAMMPS was launched with \texttt{{-}{-}cpu-bind=none}.
Per-step wall times were decomposed using \RuNNer's internal timer and cross-checked against the total step time reported by LAMMPS.
Ten molecular dynamics steps were executed per run; the first three steps were discarded to eliminate startup overhead, and the remaining seven were averaged.
All seven retained timings were required to agree to within numerical noise before a run was accepted.
A single committee member was used for all LAMMPS benchmarks.

\begin{figure}[htbp]
  \centering
  \includegraphics[width=\linewidth]{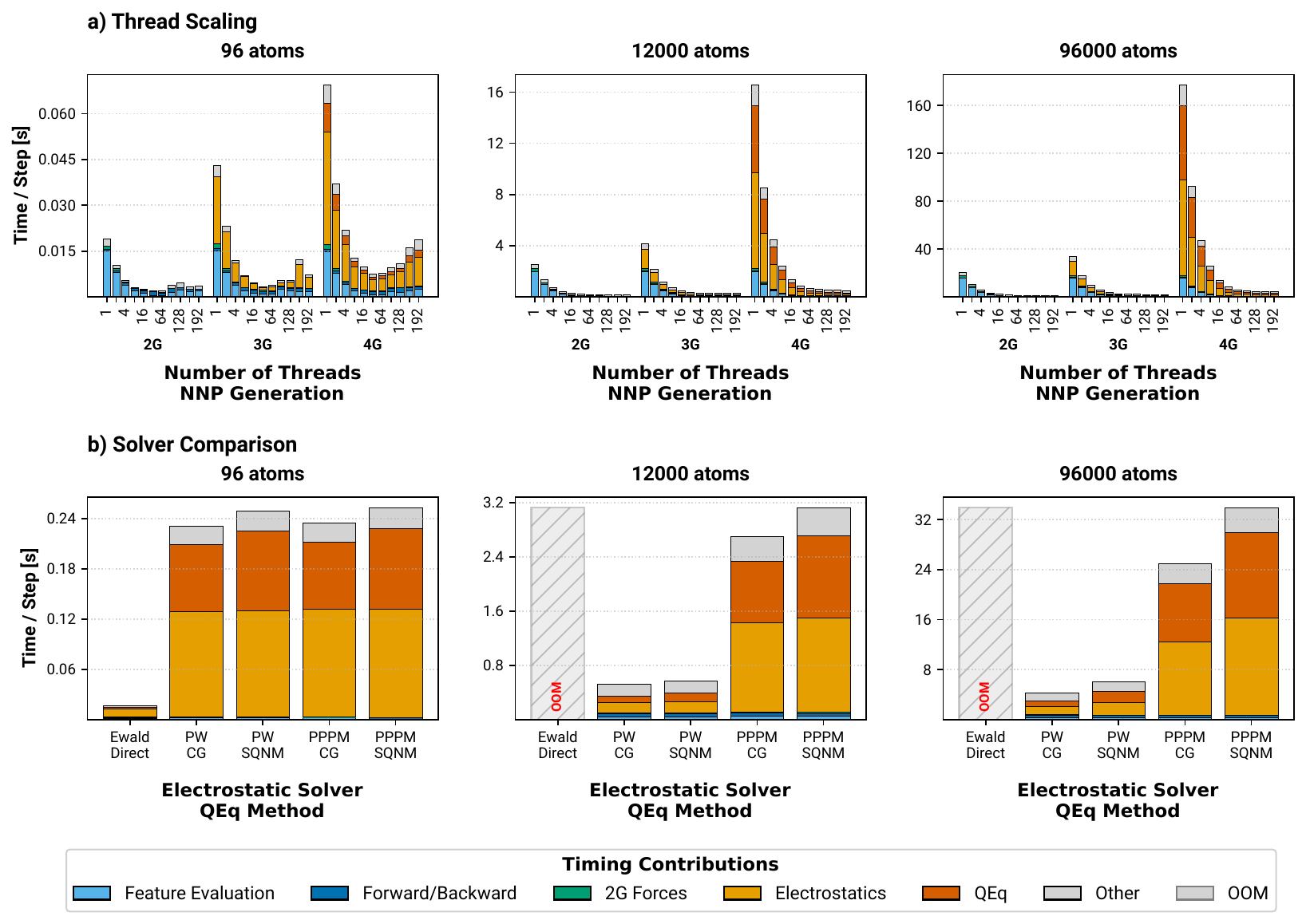}
  \caption{%
    ASE-interface prediction performance through \textsc{RuNNerlib}, decomposed into per-routine timing contributions.
    \textbf{Top row (a):} OpenMP thread scaling of the per-step wall time for the three HDNNP generations.
    \textbf{Bottom row (b):} comparison of the available electrostatic solvers (Ewald, plane-wave and PPPM variants with conjugate-gradient or SQNM charge equilibration).
    Columns correspond to the three system sizes (96, 12\,000, and 96\,000 atoms).
    Bars marked ``OOM'' indicate runs that exceeded the available memory.\\
    \vspace{2pt}
    \small\textbf{Alt Text:} Bar graphs evaluating the RuNNer 2.0 ASE interface performance across three water system sizes, showing OpenMP thread scaling by routine timing contribution and comparing execution step times across five electrostatic solver combinations.
  }
  \label{fig:si_ase_benchmarks}
\end{figure}

\paragraph{ASE interface benchmarks (Fig.~\ref{fig:ase_performance}).}
\label{sec:si:methods:perf_bechmarks:ase}
The same compilation toolchain and cluster were used for the ASE interface benchmarks.
The Python environment comprised ASE~3.28.0b1~\cite{P5915}, \textsc{runnerase-core}~0.11.1, and \textsc{runnerase-prediction}~0.3.0.
Three system sizes were tested---96, 12\,000, and 96\,000 atoms---obtained by repeating the 96-atom water unit cell as $1\times1\times1$, $5\times5\times5$, and $10\times10\times10$ supercells.
A \textsc{RuNNerlib} calculator was initialized from the \RuNNer shared library (\texttt{RuNNer.so}) located in the working directory.
Atomic velocities were assigned from a Maxwell--Boltzmann distribution at 300\,K, and the trajectory was propagated with the velocity Verlet integrator using a 0.5\,fs time step.
Wall time was measured with Python's \texttt{time.perf\_counter()} around a block of ten molecular dynamics steps, and the per-step time was obtained as the total divided by ten (no initial steps were discarded).
To probe committee scaling, runs were repeated for each combination of system size, HDNNP generation, and committee size; the committee size was varied by loading potentials comprising different numbers of the eight trained members.
In addition to the committee scaling shown in the main text (Fig.~\ref{fig:ase_performance}), the OpenMP thread scaling and the comparison of electrostatic solvers through the ASE interface are shown in Fig.~\ref{fig:si_ase_benchmarks}.

\subsection{Code Comparison Benchmarks}
\label{sec:si:methods:comparison}

The benchmarks comparing \RuNNer against n2p2~\cite{P5603} and PANNA~2.0~\cite{P6532} (Fig.~\ref{fig:lammps_performance}c--e) were performed on the Ganymede cluster at Ruhr-Universit\"at Bochum.
Each node contains two Intel Xeon Gold 6430 processors with 32 cores each (64 cores across two sockets) at up to 3.4\,GHz and 500\,GiB of RAM.

\paragraph{Comparison potential.}
\label{sec:si:methods:comparison:pot}
To maximize comparability across codes, a separate 2G- and 4G-HDNNP committee of eight members was trained on the same water dataset (Sec.~\ref{sec:si:methods:dataset}) with an adjusted descriptor set and network architecture.
The network was enlarged to three hidden layers of 30 nodes each (compared to two hidden layers of 20 nodes for the primary potential) to amplify the contribution of the neural network forward and backward passes in the overall timing.
Angular symmetry functions of type~9 were used instead of type~3, because PANNA's implementation omits the $r_{jk}$ contribution in its angular functions.
The hyperparameters of the radial symmetry functions were adjusted slightly to match PANNA's grid-based parameter generation scheme.
No 3G-HDNNP was trained for the comparison benchmarks.
Only the first committee member of both 2G and 4G models was used for benchmarking; several of the remaining 4G members diverged during training and were discarded.
The training RMSEs of the single member used are summarized in Table~\ref{tab:comp_rmse}.

\begin{table}[h]
  \centering
  \caption{%
    Training and test RMSEs of the single committee member used for the code comparison benchmarks (epoch 50 for 2G, epoch 40 for 4G charges, epoch 50 for 4G short-range).
    Units are as in Tables~\ref{tab:2g_rmse} and~\ref{tab:3g_rmse}.\\
    \vspace{2pt}
    \small\textbf{Alt Text:} Training and test root mean squared errors for the single committee member used for comparing RuNNer 2.0, n2p2, and PANNA.
  }
  \label{tab:comp_rmse}
  \begin{tabular}{lSSSS}
    \toprule
    Model
    & \multicolumn{2}{c}{Energy or charge RMSE}
    & \multicolumn{2}{c}{Force RMSE (meV\,\AA$^{-1}$)} \\
    \cmidrule(lr){2-3}\cmidrule(lr){4-5}
    & {Train} & {Test} & {Train} & {Test} \\
    \midrule
    2G (meV\,atom$^{-1}$)       & 1.4881 & 1.7162 & 167.82 & 175.37 \\
    4G charges ($m|e|$)          & 5.5767 & 6.1757 & \multicolumn{2}{c}{---} \\
    4G short (meV\,atom$^{-1}$)  & 0.4500 & 2.1812 & 252.65 & 253.45 \\
    \bottomrule
  \end{tabular}
\end{table}

\paragraph{Caveats.}
\label{sec:si:methods:comparison:caveats}
The comparison between codes is subject to several important caveats.
First, PANNA's angular symmetry functions use a slightly different functional form from those in \RuNNer and n2p2, which introduces a minor methodological non-equivalence that is favorable to PANNA's feature evaluation performance.
Second, for 4G benchmarks PANNA predicts the atomic electronegativity, hardness, and short-range energy simultaneously with a single neural network that produces multiple output nodes, in contrast to the separate network instances used by \RuNNer and n2p2.
Third, PANNA provides a GPU-enabled LAMMPS interface via Kokkos and an integration with JAX-MD; neither was available at the time of testing, so all comparison benchmarks used the CPU LAMMPS interface.
Fourth, n2p2 and \RuNNer produce near-identical energies and forces for 2G-HDNNPs; residual differences for 4G-HDNNPs are attributed to minor differences in electrostatic summation order.

\paragraph{Code modifications.}
\label{sec:si:methods:comparison:modifications}
Prior to compilation, targeted modifications were applied to PANNA and n2p2 to address correctness issues and enable optimal performance.
For PANNA, a patch file resolving a runtime crash is provided alongside the benchmark inputs at \url{https:/gitlab.com/runner-suite/runner2-paper-data}.
For n2p2, an optimized iterative QEq solver~\cite{kocer_iterative_qeq_2025} was implemented; the changes have been submitted as merge request \texttt{209} to the upstream repository.\cite{knoll_n2p2_pr_2024}
Without this fix, the n2p2 force evaluation under 4G-HDNNP exhibited cubic scaling; after the fix, the expected $\mathcal{O}(N^2)$ scaling is recovered.

\begin{figure}[htbp]
  \centering
  \includegraphics[width=\linewidth]{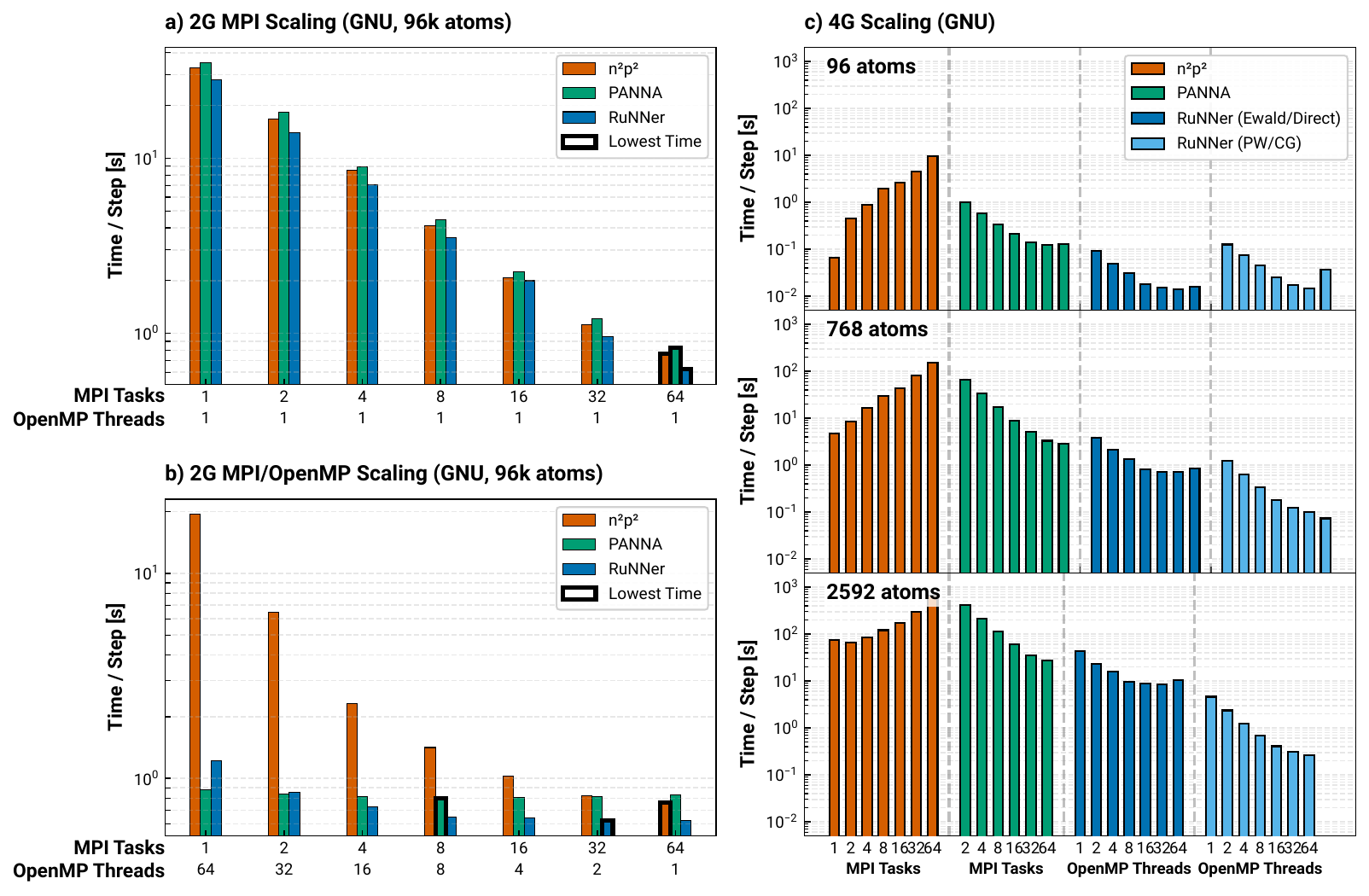}
  \caption{%
    LAMMPS prediction performance of \RuNNer, n2p2, and PANNA compiled with the GNU toolchain (GCC, OpenMPI), mirroring the Intel-toolchain results in the main text (Fig.~\ref{fig:lammps_performance}c--e).
    \textbf{Left (a, b):} 2G-HDNNP pure-MPI strong scaling (a) and mixed MPI/OpenMP scaling at a constant total thread count (b), for 96\,000 atoms.
    \textbf{Right (c):} 4G-HDNNP scaling for three system sizes (96, 768, and 2592 atoms); for \RuNNer, both the Ewald/direct and plane-wave conjugate-gradient electrostatic solvers are shown.\\
    \vspace{2pt}
    \small\textbf{Alt Text:} Bar graphs comparing execution step times for RuNNer 2.0, n2p2, and PANNA compiled with GNU tools in LAMMPS, displaying 2G pure-MPI and hybrid MPI/OpenMP scaling on 96,000 water atoms alongside 4G scaling across three system sizes.
  }
  \label{fig:si_lammps_gnu}
\end{figure}

\paragraph{Compilation.}
\label{sec:si:methods:comparison:compilation}
All codes were compiled with the Intel oneAPI toolchain: Intel Fortran/C/C++ compilers 2025.3, MKL~2025.3, and Intel MPI~2021.17.
PANNA additionally required Eigen~3.4.0 and GSL~2.8, which were compiled with the same compilers.
Optimization flags were set to \texttt{-O3} with link-time optimization and \texttt{-xCORE-AVX512 -msse4.2} to target the AVX-512 instruction set of the Xeon Gold 6430.
LAMMPS was built with the \texttt{PKG\_INTEL} package enabled.
All benchmarks were additionally repeated with a GNU-based toolchain (GCC, OpenMPI) to verify that no code was placed at an unfair disadvantage by the choice of compiler; the relative performance of all three codes was comparable, as shown for the 2G- and 4G-HDNNP comparisons in Fig.~\ref{fig:si_lammps_gnu}.
Because n2p2 and PANNA parallelize electrostatics exclusively via MPI, the MPI-parallel build was used for all codes in this benchmark set.
Inspection of the source code confirmed that while PANNA achieves excellent OpenMP parallelization for 2G evaluation, neither PANNA nor n2p2 apply OpenMP parallelization to the 4G electrostatics; both therefore rely solely on MPI task-level parallelism for 4G benchmarks.
In contrast, \RuNNer parallelizes all QEq and electrostatics operations with OpenMP, which is why the 4G OpenMP-scaling results for \RuNNer are presented separately from those of the other codes.

\paragraph{Simulation protocol.}
\label{sec:si:methods:comparison:simulation}
All code comparison benchmarks used the same system (96-atom bulk water unit cell replicated to the target system size) and the same simulation protocol as described in Sec.~\ref{sec:si:methods:perf_bechmarks}: ten NVT steps at 300\,K, first three discarded, remaining seven averaged.
Throughput is reported in femtoseconds per day as recorded by LAMMPS.
Benchmark runs were performed on exclusive nodes.
\RuNNer (version \texttt{2.0.0-alpha.11}, Git commit \texttt{ef43e5f0bf}),
n2p2 (development version, Git commit \texttt{ec1953d856}), and
PANNA (Git commit \texttt{132a2df662})
were used.

\subsection{Training Performance Benchmarks}
\label{sec:si:methods:training}

\begin{figure}[htbp]
  \centering
  \includegraphics[width=.5\linewidth]{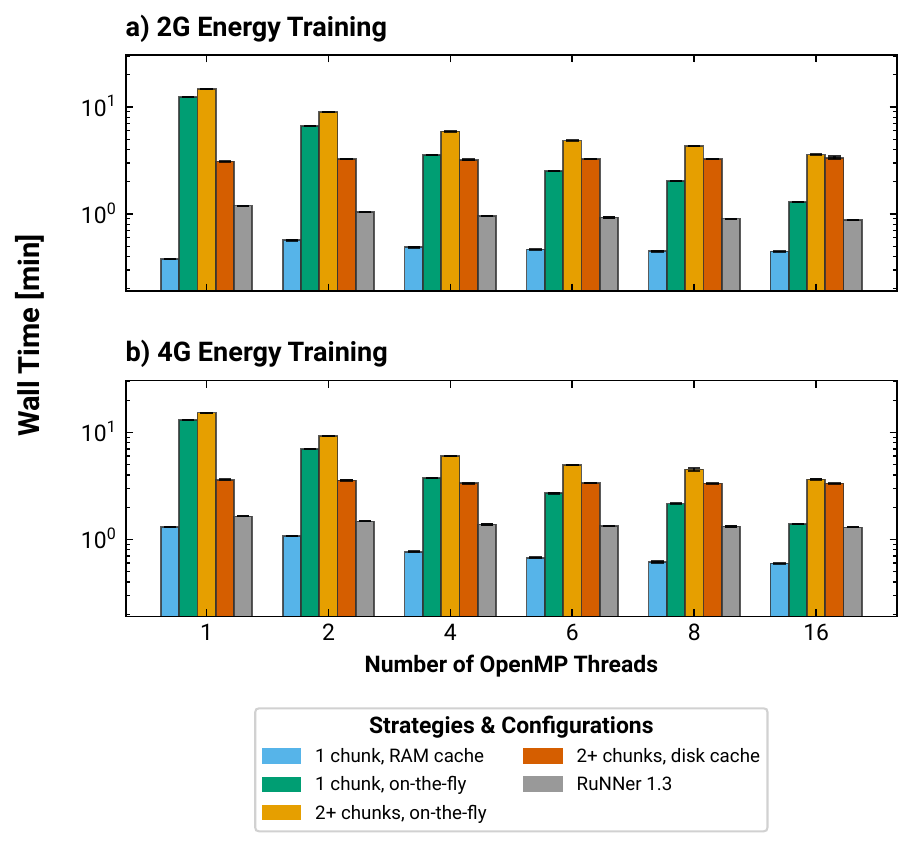}
  \caption{%
    Per-epoch wall-time comparison of \RuNNer against RuNNer~1.3 for 2G and 4G energy training as a function of the number of OpenMP threads.
    Bars are grouped by memory strategy (see legend); wall times include the precomputation step and are averaged over three independent runs.
    This figure complements the force- and charge-training benchmarks of the main text (Fig.~\ref{fig:training_performance}).\\
    \vspace{2pt}
    \small\textbf{Alt Text:} Grouped bar graphs showing training wall times per epoch for 2G and 4G energy training comparing RuNNer 2.0 across four memory management strategies against RuNNer 1.3 across one to sixteen OpenMP threads.
  }
  \label{fig:si_training_energy}
\end{figure}

Training benchmarks compare \RuNNer against RuNNer~1.3~\cite{P4444,P5128} (Git commit \texttt{d4a0edf0d5}) for 2G force training, 4G force training, and 4G charge training (Fig.~\ref{fig:training_performance}), as well as 2G and 4G energy training (Fig.~\ref{fig:si_training_energy}).

\paragraph{Dataset.}
\label{sec:si:methods:training:dataset}
The full 140-structure liquid water dataset described in Sec.~\ref{sec:si:methods:dataset} was used with the same train/test split.

\paragraph{Hardware and compilation.}
\label{sec:si:methods:training:hardware_comp}
Benchmarks were performed on the Ganymede cluster using the Intel oneAPI toolchain described in Sec.~\ref{sec:si:methods:comparison}.
RuNNer~1.3 was compiled with the identical toolchain (Intel Fortran/C/C++ 2025.3, MKL~2025.3, Intel MPI~2021.17) and the same optimization flags.
Training was managed through \textsc{runnerase-training}~\texttt{0.7.2} using the \textsc{RuNNerbin} calculator interface.
Per-epoch wall times were recorded by \RuNNer's internal timer, which measures individual computational stages (feature evaluation, forward and backward pass, optimizer step, and, for 3G/4G, electrostatics and charge equilibration). Three training runs per memory setting were averaged.

\paragraph{Training settings.}
\label{sec:si:methods:training:settings}
Since the goal of this benchmark was to provide representative timings, both codes were configured with RuNNer-typical default settings. We use two hidden layers of 15 nodes each and hyperbolic tangent activation functions.
To ensure that the two codes perform numerically identical update sequences---and thus provide a strictly controlled timing comparison---dataset shuffling was disabled and all loss thresholds were set to zero, forcing every structure and force component to be processed in the same order and at every epoch.
RuNNer~1.3 was initialized with the same Xavier--Eckhoff weight matrix as \RuNNer (Sec.~\ref{section:si:nn:act:init:xavier}), ensuring both codes begin from the same point in parameter space.
Training ran for 100 epochs to provide robust per-epoch statistics.

The optimizer in both codes was the per-element extended Kalman filter with separate covariance matrices for each element type.
This is the standard configuration in RuNNer~1.3; \RuNNer's default uses a single global covariance matrix, but the per-element variant was selected here for direct comparability.
All energies were used for training; atomic forces were included with a selection fraction of 10\,\% per epoch.
The radial cutoff radius was set to 12\,Bohr for all symmetry functions.

For 4G charge training, both the electronegativity network and the hardness network were optimized simultaneously.
The hardness output was constrained to positive values via a squared activation function on the output node.
The update scheme followed the RuNNer~1.3 default of one Kalman filter update per element type per structure.

\subsection{Non-Local Potential for AuMgO(*Al)}
\label{sec:si:methods:4G}

This section describes the reference dataset and the committee-training protocol
underlying the AuMgO(*Al) example, in which a single Au adatom on an MgO(001)
surface acquires a negative charge that is compensated by remote subsurface Al
dopants.
This charge transfer over several \AA{}ngstr\"om is a genuinely non-local effect
and thus serves as a stringent test of the fourth-generation (4G) electrostatics (Sec.~\ref{section:si:elec:4G})
against the strictly local second-generation (2G) description.

\paragraph{DFT reference dataset.}
\label{sec:si:methods:4G:dataset}
All reference calculations were performed with FHI-aims~\cite{P2189} (version \texttt{221103})
using the PBE exchange--correlation functional and the \texttt{light} species
defaults.
Scalar-relativistic effects were included at the atomic ZORA level, and all cells 
were treated with collinear spin polarization at overall charge neutrality.
The Brillouin zone was sampled with a $3 \times 3 \times 1$ Monkhorst--Pack grid,
and a Gaussian occupation smearing of width $0.1\,\mathrm{eV}$ was applied.
The electron density was mixed with a Pulay mixer (history of 10 steps,
mixing parameter 0.1).
Self-consistency was converged to $10^{-6}\,\mathrm{eV}$ in the total energy,
$10^{-3}\,\mathrm{eV}$ in the sum of eigenvalues, $10^{-5}\,e\,\mathrm{Bohr}^{-3}$
in the electron density, and $10^{-4}\,\mathrm{eV}\,\text{\AA}^{-1}$ in the forces.
Hirshfeld charges were extracted for 4G training purposes.

The reference structures were MgO(001) slabs built from the bulk lattice
constant $a \approx 4.261\,\text{\AA}$ as $3 \times 3$ lateral supercells
at three slab thicknesses of 3, 4, and 5 MgO layers, corresponding to
$3\times3\times3$, $3\times3\times4$, and $3\times3\times5$ cells.
A vacuum region of approximately $15\,\text{\AA}$ separated the periodic images
along the surface normal.
For the doped cells, three Mg$^{2+}$ ions in the deep subsurface layers, far
from the adsorption site, were replaced by Al$^{3+}$; charge neutrality of the cell
then drives a compensating negative charge onto the Au adatom.
A single Au atom was placed on top of the surface and sampled on a dense grid
in the wedge of non-symmetry equivalent positions and as a function of the
vertical Au--surface distance from about $1.5$ to $3.5\,\text{\AA}$. This includes
the favorable adsorption sites atop Mg or O.
For the doped $3\times3\times5$ slab, a substantial fraction of the single-point
calculations did not reach SCF convergence and are absent from the reference set;
the pristine $3\times3\times5$ slab and both thinner slabs are complete. Additionally, we include the six slabs without an adsorbed gold atom and a single gold atom in vacuum in the training dataset.

\paragraph{HDNNP committee training.}
\label{sec:si:methods:4G:training}
On this common dataset, comprising all three thicknesses, both adsorption sites,
and pristine and doped cells, a 2G-HDNNP and a 4G-HDNNP committee were trained,
each with three independently initialized members.
As the example targets the energy landscape of the adsorption process, both
committees were trained on energies only.
Both descriptions use element-embracing atom-centered symmetry
functions of radial and angular type (Sec.~\ref{section:si:features:acsf})
built for the elements O, Mg, Al, and Au, with a cosine cutoff of $10.0\,\mathrm{Bohr}$.

The 2G-HDNNP committee employs a single short-range atomic energy network per
element with three hidden layers of 30, 25, and 20 nodes and hyperbolic-tangent activations.
Each member was trained with the Kalman-filter optimizer (Sec.~\ref{section:si:training:opt:kalman}) minimizing a mean-absolute-error loss, using a $10\,\%$ test set.

The 4G-HDNNP committee was trained in the standard two-stage 4G
protocol (Sec.~\ref{section:si:training}).
In the first stage, the environment-dependent electronegativity network
(two hidden layers of 15 nodes each, hyperbolic-tangent activations) was
trained together with trainable elemental hardnesses; the global atomic charges
follow from charge equilibration, with the electrostatic energy evaluated by
Ewald summation (Sec.~\ref{section:si:elec:ewald}) at a precision of $10^{-6}$.
Element-specific fixed Gaussian charge widths of $4.289$, $2.872$, $3.477$, and
$3.288\,\mathrm{Bohr}$ were used for O, Mg, Al, and Au, respectively.
This stage was trained for 30 epochs with the Kalman filter minimizing a
mean-squared-error loss on the reference charges, with the optimal model
selected on the test-set charge error.

In the second stage, the short-range energy network (three hidden layers of
30, 25, and 20 nodes, hyperbolic-tangent activations) was trained with the Kalman
filter and a mean-absolute-error loss. Electrostatic energies were not removed
from the DFT reference before training.
For both committees, committee uncertainties were obtained as the standard
deviation across the three members.

The two short-range energy fits (the 2G potential and the second stage of the
4G potential) were stopped manually; for both, the committee members
at the $30$th epoch were retained, as later epochs did not yield a systematic
improvement of the test error. The 4G electronegativity/charge fit was taken at
its final ($50$th) epoch. The corresponding committee-averaged root-mean-square errors (RMSEs),
averaged over the three members, are summarized in Table~\ref{tab:si:aumgo_rmse}.

\begin{table}[htbp]
\centering
\caption{Committee-averaged training and test RMSEs of the AuMgO(*Al) potentials, averaged over the three committee members. The short-range energy fits are reported at the $30$th epoch and the 4G charge fit at its final ($50$th) epoch (see text).}
\label{tab:si:aumgo_rmse}
\begin{tabular}{llcc}
\toprule
Fit & Quantity & Train & Test \\
\midrule
2G short-range        & Energy (meV/atom) & $1.69$ & $1.78$ \\
4G short-range        & Energy (meV/atom) & $0.22$ & $0.27$ \\
4G electronegativity  & Charge (m$e$)    & $4.83$  & $4.75$  \\
\bottomrule
\end{tabular}
\end{table}

\section{Use of generative AI}
During the preparation of this work the authors used Anthropic's Claude in order to generate analysis and visualization scripts, and to improve wording. After using this tool, the authors reviewed and edited the content as needed and take full responsibility for the content of the published article.

\bibliography{literature}

\end{document}